\documentclass[12pt]{article}
\usepackage{epsfig,cite}
\usepackage {amsmath}
\usepackage{amssymb}
\include{epsf}
\newcount\timecount
\newcount\hours \newcount\minutes  \newcount\temp \newcount\pmhours
\hours = \time
\divide\hours by 60
\temp = \hours
\multiply\temp by 60
\minutes = \time
\advance\minutes by -\temp
\def\hour{\the\hours}
\def\minute{\ifnum\minutes<10 0\the\minutes
            \else\the\minutes\fi}
\def\clock{
\ifnum\hours=0 12:\minute\ AM
\else\ifnum\hours<12 \hour:\minute\ AM
      \else\ifnum\hours=12 12:\minute\ PM
            \else\ifnum\hours>12
                 \pmhours=\hours
                 \advance\pmhours by -12
                 \the\pmhours:\minute\ PM
                 \fi
            \fi
      \fi
\fi
}

\def\monthname{\relax\ifcase\month 0/\or January\or February\or
   March\or April\or May\or June\or July\or August\or September\or
   October\or November\or December\else\number\month/\fi}

\def\bold#1{\setbox0=\hbox{$#1$}%
     \kern-.025em\copy0\kern-\wd0
     \kern.05em\copy0\kern-\wd0
     \kern-.025em\raise.0433em\box0 }

\def\beq{\begin{equation}}
\def\eeq{\end{equation}}

\def\gev{{\rm \, Ge\kern-0.125em V}}
\def\tev{{\rm \, Te\kern-0.125em V}}
\def\gyr{{\rm \, G\kern-0.125em yr}}
\def\ohsq{\Omega_{\chi} h^2}

\def\nl{\hfill\nonumber\\&&}
\def\nnl{\hfill\nonumber\\}

\def\gappeq{\gtrsim}
\def\lappeq{\lesssim}
\def\Toprel#1\over#2{\mathrel{\mathop{#2}\limits^{#1}}}

\def\stau{\widetilde \tau}

\def\m12{m_{1\!/2}}

\def\PL{{Phys.~Lett.} }
\def\PR{{Phys.~Rev.} }

\def\stau{\tilde{\tau}}

\def\mpl{M_{P}}

\def\bea{\begin{eqnarray}}
\def\eea{\end{eqnarray}}

\def\mplr{\overline{M_{P}}}
\def\mgut{M_{GUT}}
\def\calh{\mathcal{H}}
\def\BSig{B_{\Sigma}}
\def\mSig{m_{\Sigma}}
\def\muS{\mu_{\Sigma}}
\def\hfiv{h_\mathbf{\overline{5}}}
\def\hten{h_\mathbf{10}}
\def\mfiv{m_\mathbf{\overline{5}}}
\def\mfivl{m_{\mathbf{\overline{5}},1}}
\def\mten{m_\mathbf{10}}
\def\mtenl{m_{\mathbf{10},1}}
\def\afiv{A_\mathbf{\overline{5}}}
\def\aten{A_\mathbf{10}}
\def\alam{A_{\lambda}}
\def\alamp{A_{\lambda'}}

\mathchardef\mhyphen="2D

\DeclareMathOperator{\Tr}{Tr}

\begin{document}
\begin{titlepage}
\pagestyle{empty}
\baselineskip=21pt
\rightline{CERN-PH-TH/2010-080}
\rightline{UMN--TH--2902/10}
\rightline{FTPI--MINN--10/11}
\vskip 0.2in
\begin{center}
{\large{\bf Resurrecting No-Scale Supergravity Phenomenology}}
\end{center}
\begin{center}
\vskip 0.2in
{\bf John~Ellis}$^1$, {\bf Azar Mustafayev}$^{2}$ and {\bf Keith~A.~Olive}$^{2}$

\vskip 0.1in

{\it
$^1${TH Division, PH Department, CERN, CH-1211 Geneva 23, Switzerland}\\
$^2${William I.~Fine Theoretical Physics Institute, \\
University of Minnesota, Minneapolis, MN 55455, USA}\\
}

\vskip 0.2in
{\bf Abstract}
\end{center}
\baselineskip=18pt \noindent

In the context of phenomenological models in which the soft supersymmetry-breaking
parameters of the MSSM become universal
at some unification scale, $M_{in}$, above the GUT scale, $\mgut$, it is possible
that all the scalar mass parameters $m_0$, the trilinear couplings $A_0$ and the bilinear
Higgs coupling $B_0$ vanish simultaneously, as in no-scale supergravity.  Using these
no-scale inputs in a renormalization-group analysis of the minimal supersymmetric SU(5)
GUT model, we pay careful
attention to the matching of parameters at the GUT scale. We delineate the region
of $M_{in}$, $m_{1/2}$ and $\tan \beta$ where the resurrection of no-scale
supergravity is possible, taking due account of the relevant phenomenological
constraints such as electroweak symmetry breaking, $m_h, b \to s \gamma$, the 
neutralino cold
dark matter density $\ohsq$ and  $g_\mu - 2$. No-scale supergravity
survives in an L-shaped strip of parameter space, with one side having with one side having
$m_{1/2} \gappeq 200~{\rm GeV}$, the second (orthogonal) 
side having  $M_{in} \gappeq 5 \times 10^{16}$~GeV. 
Depending on the relative signs and magnitudes of the GUT superpotential
couplings, these may be connected to form a
triangle whose third side is a hypotenuse at larger $M_{in}$, $m_{1/2}$ and $\tan \beta$,
whose presence and location depend on the GUT superpotential parameters. We
compare the prospects for detecting sparticles at the LHC in no-scale supergravity 
with those in the CMSSM and the NUHM.


\vfill
\leftline{CERN-PH-TH/2010-080}
\leftline{April 2010}
\end{titlepage}

\section{Introduction}

Many of the uncertainties in the low-energy phenomenology of supersymmetry
are associated with the pattern of supersymmetry breaking. Phenomenologists
usually take a {\it bottom-up} approach, {\it e.g.}, postulating some form of universality
for the soft supersymmetry-breaking (SSB) parameters $m_0, m_{1/2}$ and $A_0$,
motivated by the observed suppression of flavour-changing neutral interactions~\cite{EN}.
This suggests that, before renormalization, the SSB scalar masses 
are (approximately) universal for squarks and sleptons in different generations that
have the same electroweak quantum numbers~\cite{BG}. Grand Unified Theories (GUTs)
such as SU(5) further suggest universality between the 
the SSB scalar masses 
for squarks and sleptons in the same GUT multiplet, {\it e.g.}, the left-handed sleptons and the
right-handed charge-$(-1/3)$ squarks in a fiveplet of SU(5). Universality between the
the SSB scalar masses 
in the fiveplet and tenplet of SU(5) might be motivated by
a larger GUT such as SO(10), but universality with the 
SSB 
masses of the Standard Model Higgs multiplets would require some principle beyond
GUTs. Nevertheless, this extended universality is often postulated,
particularly at the GUT scale in the framework of
the constrained MSSM (CMSSM)~\cite{funnel,cmssm,efgosi,cmssmnew,cmssmmap,like1,like2}, 
though models with non-universal Higgs masses (NUHM) 
are also often studied~\cite{nonu,nuhm,like2}. All of these models are usually analyzed assuming that the gravitino
is heavy and the lightest supersymmetric particle (LSP) is the lightest 
neutralino $\chi$~\cite{EHNOS}.

On the other hand, taking a {\it top-down} point of view,
supersymmetry breaking presumably
originates from the spontaneous breaking of local supersymmetry via the
super-Higgs mechanism in $N = 1$ supergravity~\cite{Polonyi,Fetal,BIM}. Therefore the pattern of
supersymmetry breaking is linked to the form of the effective supergravity theory that, in
turn, presumably originates from some form of superstring theory.
One of the most popular hypotheses is that of minimal supergravity (mSUGRA)
in which the K\"ahler potential $K$ characterizing the kinetic terms of the
chiral supermultiplets $\Phi_i$ has the simple form $K = \sum_i |\Phi_i|^2$~\cite{bfs}. In this
case, universality of the $m_0$ and $A_0$ would be automatic and there would, in
addition, be a relation for the soft bilinear Higgs coupling $B_0 = A_0 - m_0$ 
and the gravitino mass would be fixed: $m_{3/2} = m_0$~\cite{bfs,vcmssm}. 
The first of these constraints can be regarded as fixing, via
the electroweak vacuum conditions~\cite{rewsb}, $\tan \beta$ as a function of $m_{1/2}, m_0$
and $A_0$, and the latter condition implies that the gravitino is the LSP in
extensive regions of parameter space. Hence mSUGRA is not equivalent to the CMSSM.

Other forms of effective supergravity theory have also been suggested, and the
possibility we study here is no-scale supergravity~\cite{nosc1}, in which $m_0 = A_0 = B_0 = 0$
before renormalization~\footnote{The gravitino is not expected to be the LSP in such
no-scale models.}. The K\"ahler potential for the simplest form of no-scale
supergravity for a single chiral supermultiplet $T$ is $K = - 3 {\rm ln} (T + T^\dagger)$,
and the simplest form for many additional chiral supermultiplets $\Phi_i$ is 
$K = - 3 {\rm ln} (T + T^\dagger - \sum_i |\Phi_i|^2)$. Many other forms of K\"ahler potential
also have the no-scale property, notably those derived as effective theories
from string models~\cite{Witten}, where $T$ can be regarded as a prototype modulus field.
We note that the no-scale boundary conditions also arise in models of 
gaugino mediation~\cite{gaugino} considered in the context of brane-world models.

However, for many years no-scale supergravity phenomenology has been disfavoured, because data
seemed to disfavour small values of $m_0$ and specifically $m_0 = 0$. For example,
in~\cite{ENO} it was concluded that  a framework with unified gaugino masses and 
$m_0 = 0$ at the GUT scale would be possible only for a very
restricted range of $\tan \beta \sim 8$~\footnote{One can remove the 
unpleasant stau LSP feature by considering non-universal gaugino masses arising from
a restricted gauge symmetry on the hidden brane~\cite{Baer:2002by}.}. 
It was also suggested~\cite{ENO} that
no-scale supergravity could be rescued over a larger range of $\tan \beta$ if the
no-scale boundary condition applied at the Planck scale.
Indeed, it was recognized in Ref.~\cite{SS} that the problem of a stau LSP can be
alleviated when the unification scale is raised sufficiently above the GUT scale,
leaving the possibility open for a bino LSP as dark matter. 
The superpartner spectrum
found in gaugino mediation models  with a high unification scale was discussed in Ref.~\cite{SS2}.

We have recently studied~\cite{EMO} how the parameter space of the CMSSM
is modified if 
SSB scalar mass
universality is imposed at some scale $M_{in} > \mgut$.
Specifically, we studied the minimal SU(5) GUT model~\cite{pp} 
with $m_0$ defined  
at some scale $\mgut \le M_{in} \le \mplr \equiv \mpl /\sqrt{2\pi} \sim 2.4 \times 10^{18}$~GeV~\footnote{For 
a recent review of this sample model and its compatibility with experiment, 
see~\cite{Senjanovic:2009kr}.}.
The regions of the $(m_{1/2}, m_0)$ plane favoured by the available
phenomenological constraints such as the density of neutralino dark matter $\ohsq$ change
substantially with $M_{in}$ and we confirmed, in particular, the previous observations~\cite{ENO,SS}
that the option $m_0 = 0$ could be permitted under some circumstances. 
This study did not, however, apply directly to no-scale models, since the conditions
$A_0 = B_0 = 0$ were not imposed at $M_{in}$.

In this paper we study systematically the consequences of applying the full
no-scale conditions $m_0 = A_0 = B_0 = 0$ at some common scale $M_{in} > M_{GUT}$
in the framework of the simplest SU(5) GUT. Assuming that the gravitino is heavy,
the only relevant supersymmetry-breaking parameter in this model is $m_{1/2}$.
However, in principle, even the simplest SU(5) also has two GUT superpotential parameters
$\lambda, \lambda'$~\cite{EMO}, 
whose values influence the allowed region in the no-scale $(m_{1/2}, M_{in})$
plane. We find here that
the region of this plane allowed by all the phenomenological constraints is a narrow 
WMAP-compatible L-shaped strip with a rounded corner. Its near-vertical part has
$m_{1/2} \gappeq 200~{\rm GeV}$, and the near-horizontal second
side has  $M_{in} \gappeq 5 \times 10^{16}$~GeV. Depending on the magnitudes
and relative signs of the GUT 
superpotential parameters, the L shape may become a triangle whose
third side (the hypotenuse) connects the ends of the L shape though larger values of
$M_{in}$, $m_{1/2}$ and $\tan \beta$.
The triangle contracts to a `blob' and then disappears as $- \lambda$ increases,
for any fixed value of $\lambda' > 0$. Based on this analysis, we then discuss the prospects
for detecting supersymmetry at the LHC within the no-scale supergravity framework.

\section{The Minimal SU(5) GUT Superpotential and RGEs}
\label{sec:su5}

In order to discuss the evolution of model parameters above the GUT scale,
we adopt the minimal SU(5) GUT model~\cite{pp} with the conventional assignments of
matter superfields to $\bf{\overline{5}}$ and $\bf{10}$ representations.
In this model, SU(5) is broken down to the Standard Model gauge
group by a single adjoint Higgs multiplet $\hat{\Sigma}(\bf{24})$, 
and the renormalizable part of the superpotential for this and the 
two five-dimensional SU(5) representations $\hat{\calh}_1(\bf{\overline{5}})$ and $\hat{\calh}_2(\bf{5})$ is
\beq
W_H = \mu_\Sigma \Tr\hat{\Sigma}^2 + \frac{1}{6}\lambda'\Tr\hat{\Sigma}^3
+ \mu_H \hat{\calh}_{1\alpha} \hat{\calh}_2^{\alpha} 
+ \lambda \hat{\calh}_{1\alpha} \hat{\Sigma}^{\alpha}_{\beta} \hat{\calh}_2^{\beta} \ ,
\label{WH}
\eeq
where Greek letters denote SU(5) indices. 
The corresponding soft SUSY-breaking lagrangian terms are
\bea
\mathcal{L}_{soft}&\ni& -m^2_{\calh_1}|\calh_1|^2 -m^2_{\calh_2}|\calh_2|^2 
                      -m^2_{\Sigma}\Tr(\Sigma^\dagger \Sigma) \nl
		      -\left[\BSig\muS \Tr\Sigma^2 +\frac{1}{6}\alamp\lambda'\Tr\Sigma^3
		      +B_H\mu_H \calh_{1\alpha}\calh_2^{\alpha} 
		      +\alam\lambda\calh_{1\alpha}\Sigma^{\alpha}_{\beta}\calh_2^{\beta} +h.c. \right] .
\label{softH}
\eea
In addition, $\mathcal{L}_{soft}$ also contains mass terms for the gaugino fields ($M_5$) for the
first- and second-generation fermionic fields ($\mfivl$ and $\mtenl$) 
and their third-generation counterparts ($\mfiv$ and $\mten$), as well as trilinear scalar couplings 
($\afiv$
and $\aten$). Note that $\mu_H$ and $\muS$ are of $\mathcal{O}(\mgut)$, while the rest of the soft parameters
are of $\mathcal{O}(M_{weak})$.

The minimal SU(5) GUT model assumes universality of corresponding soft SUSY-breaking terms and 
is completely specified the following set of parameters
\beq
m_0,\ m_{1/2},\ A_0,\ B_0,\ M_{in},\ \lambda,\ \lambda',\ sgn(\mu).
\eeq
In its no-scale incarnation, we impose at the scale $M_{in}$
\bea
 \mfivl=\mtenl=\mfiv=\mten=m_{\calh_1}=m_{\calh_2}=m_{\Sigma} &\equiv & m_0 = 0, \nnl
 \afiv=\aten=A_{\lambda}=A_{\lambda'} &\equiv & A_0 = 0, \nnl
 \BSig = B_H & \equiv  & B_0 = 0, \nnl
M_5 &\equiv & m_{1/2} \ .
\label{BC1}
\eea
These soft parameters along with gauge and Yukawa couplings are evolved between $M_{in}$ and $\mgut$ 
using the SU(5) RGEs given in Ref.~\cite{pp}.

The renormalization-group equations (RGEs)
for the third-generation matter Yukawa couplings $\hfiv$ and $\hten$ 
between $M_{in}$ and $\mgut$ are:
\bea
 \frac{d \hfiv}{dt} & = & 
  \frac{\hfiv}{16 \pi^2} \left[ 5\hfiv^2 + 48\hten^2 + \frac{24}{5} \lambda^2 
  - \frac{84}{5} g_5^2 \right], \label{RGEhfiv} \\
 \frac{d \hten}{dt} & = & 
  \frac{\hten}{16 \pi^2} \left[ 144\hten^2 + 2\hfiv^2 + \frac{24}{5} \lambda^2 
  - \frac{96}{5} g_5^2 \right],
\label{RGEhten}
\eea
where $g_5$ is the SU(5) gauge coupling above the GUT scale.
We note that the Yukawa coupling $\lambda$, but not $\lambda'$, contributes directly to the
RGEs for $\hfiv$ and $\hten$. Likewise, the RGEs for the most relevant trilinear parameters 
between $M_{in}$ and $\mgut$ also involve $\lambda$ but not $\lambda'$:
\bea
 \frac{d \afiv}{dt} & = & 
  \frac{1}{8 \pi^2} \left[ 5\afiv\hfiv^2 +48\aten\hten^2 +\frac{24}{5}\alam\lambda^2
  -\frac{84}{5} g_5^2 M_5 \right],  \label{RGEAt}\\
 \frac{d \aten}{dt} & = & 
  \frac{1}{8 \pi^2} \left[ 2\afiv\hfiv^2 +144\aten\hten^2 +\frac{24}{5}\alam\lambda^2
  -\frac{96}{5} g_5^2 M_5 \right], \label{RGEAb}
\eea
as do the RGEs for the soft supersymmetry-breaking 
squared scalar masses of the electroweak Higgs multiplets.
The appearances of $\lambda$ in these RGEs imply that our results are more
sensitive to this coupling than to $\lambda'$, as we discuss later.

The matching to MSSM parameters is done at $\mgut$ as following:
\bea
g_1=g_2=g_3 = g_5 \, , && h_t =4\hten \, ,\nnl
M_1=M_2=M_3 = M_5 \, , && (h_b+h_\tau)/2=\hfiv/\sqrt{2} \, , \nnl
A_t = \aten \, , && A_b =A_{\tau} =\afiv \, , \nnl
m_{D_1}^2 = m_{L_1}^2 = \mfivl^2 \, , && m_{Q_1}^2=m_{U_1}^2=m_{E_1}^2 = \mtenl^2 \, , \nnl
m_{D_3}^2 = m_{L_3}^2 = \mfiv^2 \, , && m_{Q_3}^2=m_{U_3}^2=m_{E_3}^2 = \mten^2 \, , \nnl
m_{H_d}^2 = m_{\calh_1}^2 \, , && m_{H_u}^2 = m_{\calh_2}^2 \, .
\label{matching1}
\eea
Below $\mgut$, the standard MSSM RGEs~\cite{Martin:1993zk} are used to obtain values of soft parameters at
lower scales~\footnote{For the MSSM RGEs, we use the convention of \cite{dlt} which 
have opposite signs for all trilinear and bilinear SSB terms relative to those in \cite{Martin:1993zk}.}. 
As mentioned in Ref~\cite{EMO}, we do not impose the condition of 
exact $b-\tau$ yukawa unification at $\mgut$,
since there could be non-renormalizable operators in SU(5) that are necessary to correct the poor Yukawa relations
for the lighter families.

The Higgs bilinear superpotential $\mu$ terms and the soft 
supersymmetry-breaking $B$ terms decouple from the rest of RGEs. This
enables one to use the EWSB minimization conditions~\cite{rewsb} to trade $\tan\beta$
for $B_0$ and predict $\mu^2$ as a function of soft parameters and $B_0$.
In the true no-scale framework, the $B$ terms must also vanish at $M_{in}$, as
seen in (\ref{BC1})~\footnote{As we will see later, this condition imposes very strong 
constraints on the parameter space and 
the allowed dark-matter annihilation mechanism, as compared to minimal SU(5) with vanishing 
$m_0$ and $A_0$, but arbitrary
$\BSig  ,\ B_H$, that was discussed for example in Ref.~\cite{EMO,quasinoscale}.}.
The $B$ terms are then evolved down to $\mgut$ using the following RGEs:
\bea
 \frac{d B_H}{dt} & = & \frac{1}{8 \pi^2} 
   \left[ 48\aten\hten^2 +2\afiv\hfiv^2 +\frac{48}{5}\alam\lambda^2 
          -\frac{48}{5}M_5 g_5^2 \right]  , \nonumber\\
 \frac{d \BSig}{dt} & = & \frac{1}{8 \pi^2} 
   \left[ 2\alam\lambda^2 +\frac{21}{10}\alamp\lambda'^2 -20M_5 g_5^2 \right] .
\label{RGEb}
\eea
In this case, $\tan \beta$ is fixed as a function of the other parameters by the electroweak vacuum 
conditions, and the free parameters of the model are simply
\beq
m_{1/2},\ M_{in},\ \lambda,\ \lambda'
\label{param}
\eeq
and, motivated by $g_\mu - 2$ measurements~\cite{newBNL,g-2}, we choose $\mu > 0$. 

It was observed in~\cite{vcmssm} that the value of the MSSM parameter $B$ at the GUT scale has a strong
influence in low-energy parameters, particularly $\tan \beta$. Accordingly, we pay close
attention to the boundary condition for $B$ that matches correctly the GUT
renormalization downwards from $M_{in}$, where we assume $B_0 = A_0 = m_0 = 0$,
to the continuing MSSM RGE analysis down to the electroweak scale. This matching has been
studied carefully in~\cite{Borzumati}, and we present below the major steps of the derivation using our
notation.

The adjoint Higgs multiplet $\hat{\Sigma}$ can be represented by a traceless matrix:
\beq
\hat{\Sigma}^{\alpha}_{\beta} =\sqrt{2}\hat{\Sigma}_r (T_r)^{\alpha}_{\beta}  \ ,
\eeq
where the $T_r \ (r=1..24)$ are SU(5) generators with $\Tr (T_r T_s)=\delta_{rs}/2$,
The breaking $SU(5)\rightarrow SU(3)_c \times SU(2)_L \times U(1)_Y$ arises 
from the Standard-Model singlet component $\hat{\Sigma}_{24}$,
that develops a vev of $\mathcal{O}(\mgut)$, 
$\langle \hat{\Sigma}\rangle = \langle \hat{\Sigma}_{24}\rangle \, diag(2,2,2,-3,-3)$. 
The latter can be decomposed as
\beq
\langle \hat{\Sigma}_{24} \rangle = 
  \langle \Sigma_{24} \rangle + \theta^2 \langle \mathcal{F}_{24} \rangle ,
\eeq
where $\Sigma_{24}$ and $\mathcal{F}_{24}$ are, respectively, the scalar and auxiliary field components of 
superfield $\hat{\Sigma}_{24}$. 
The auxiliary component is determined from the superpotential (\ref{WH}) to be
\beq
\mathcal{F}_{24}^\dagger
  = \left(\frac{\partial W_H}{\partial \hat{\Sigma}_{24}}\right)_{\hat{\Sigma}=\Sigma}
  = 2\muS\Sigma_{24} -\frac{1}{2\sqrt{30}}\lambda' \Sigma_{24}^2  .
\eeq
We can find both the scalar and auxiliary component vevs by minimizing the relevant part of the 
scalar potential that breaks SU(5) 
\beq
 V_{\Sigma_{24}} =  |\mathcal{F}_{24}|^2 +\widetilde{V}_{\Sigma_{24}} ,
\eeq
where $\widetilde{V}_{\Sigma_{24}}$ is a subset of soft supersymmetry-breaking 
terms (\ref{softH}) that contains only $\Sigma_{24}$ fields.
Using $\delta \equiv (M_{SUSY}/\mgut)$ as an expansion parameter, 
we can find perturbative solutions of the form
\bea
 \langle\Sigma_{24}\rangle &=&  v_{24}+\delta v_{24} +\delta^2 v_{24} +
                                \mathcal{O}\left(\frac{M_{SUSY}^3}{\mgut^2}\right)  \nnl
 \langle\mathcal{F}_{24}\rangle &=&  0+ F_{24} +\delta F_{24} +
                                \mathcal{O}\left(\frac{M_{SUSY}^3}{\mgut}\right) .
\label{vevexpan}
\eea
The first terms on the right-hand sides of (\ref{vevexpan}) correspond to the case of exact supersymmetry:
since $\langle \hat{\Sigma}_{24} \rangle$ breaks only SU(5) and not supersymmetry, 
it vanishes for the auxiliary field. For the scalar component vev we get the familiar expression 
$v_{24} = 2\sqrt{30}\muS /\lambda'$. 
The subsequent terms represent corrections induced by the presence of the 
soft terms~\cite{Hall:1983iz}.

The MSSM Higgs bilinears $\mu$ and $B$ can be expressed in terms of SU(5) parameters as
\bea
\mu  &=& \mu_H -\frac{3}{\sqrt{30}}\lambda \langle\Sigma_{24}\rangle , \nnl
B\mu &=& B_H\mu_H -\frac{3}{\sqrt{30}}\lambda 
        \left( \alam \langle\Sigma_{24}\rangle + \langle\mathcal{F}_{24}\rangle \right) .
        \label{match}
\eea
Using the expansions (\ref{vevexpan}) and eliminating $\mu_H$, we obtain the following
expression for the MSSM parameter $B$ in terms of the SU(5) quantities and $\mu$:
\beq
B=B_H-\frac{6\lambda}{\mu\lambda'}\left[ (\BSig -\alamp)(2\BSig -\alamp)+\mSig^2 \right] ,
\label{Bmatch}
\eeq
where we have used the fact that the combination $\Delta\equiv B_H-\alam-\BSig-\alamp$ 
is a RGE invariant (at one loop) and therefore remains 
zero at all scales. To help understand the behaviour of this matching condition
we also list the relevant RGEs for $\alamp$ and $\mSig^2$,
\bea
 \frac{d \alamp}{dt} & = & 
  \frac{1}{8 \pi^2} \left[ 3\lambda^2 \alam +\frac{63}{20}\lambda'^2\alamp
  -30 g_5^2 M_5 \right],  \\
 \frac{d \mSig^2}{dt} & = & 
  \frac{1}{8 \pi^2} \left[ \lambda^2 \left( m_{\calh_1}^2 +m_{\calh_2}^2 +\mSig^2 +\alam^2 \right)
  +\frac{21}{20}\lambda'^2 \left( 3\mSig^2 +\alamp^2 \right)
  -20 g_5^2 M_5^2 \right]. 
  \label{morerge}
\eea
We note in passing that all the RGEs shown above involve just the squares of $\lambda$
and $\lambda'$, and hence are insensitive to their signs. However, the relative sign of
$\lambda$ and $\lambda'$ does enter into the matching conditions (\ref{match}), 
with the consequences for phenomenology that we discuss in the next Section.

\section{The $(m_{1/2}, M_{in})$ Plane of No-Scale Supergravity}
\label{sec:plane}

\begin{figure}
\begin{center}
\epsfig{file=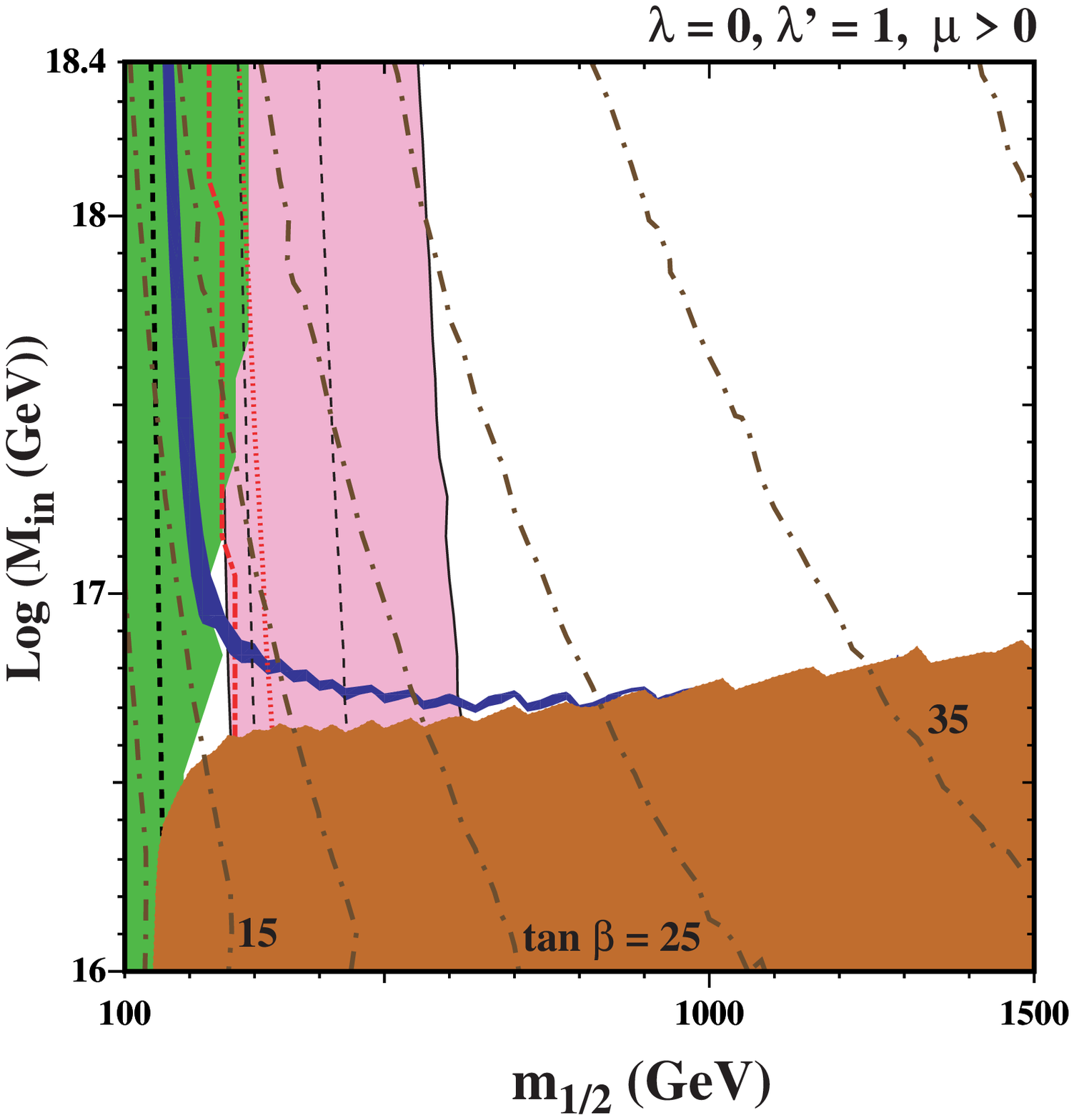,height=8cm}
\epsfig{file=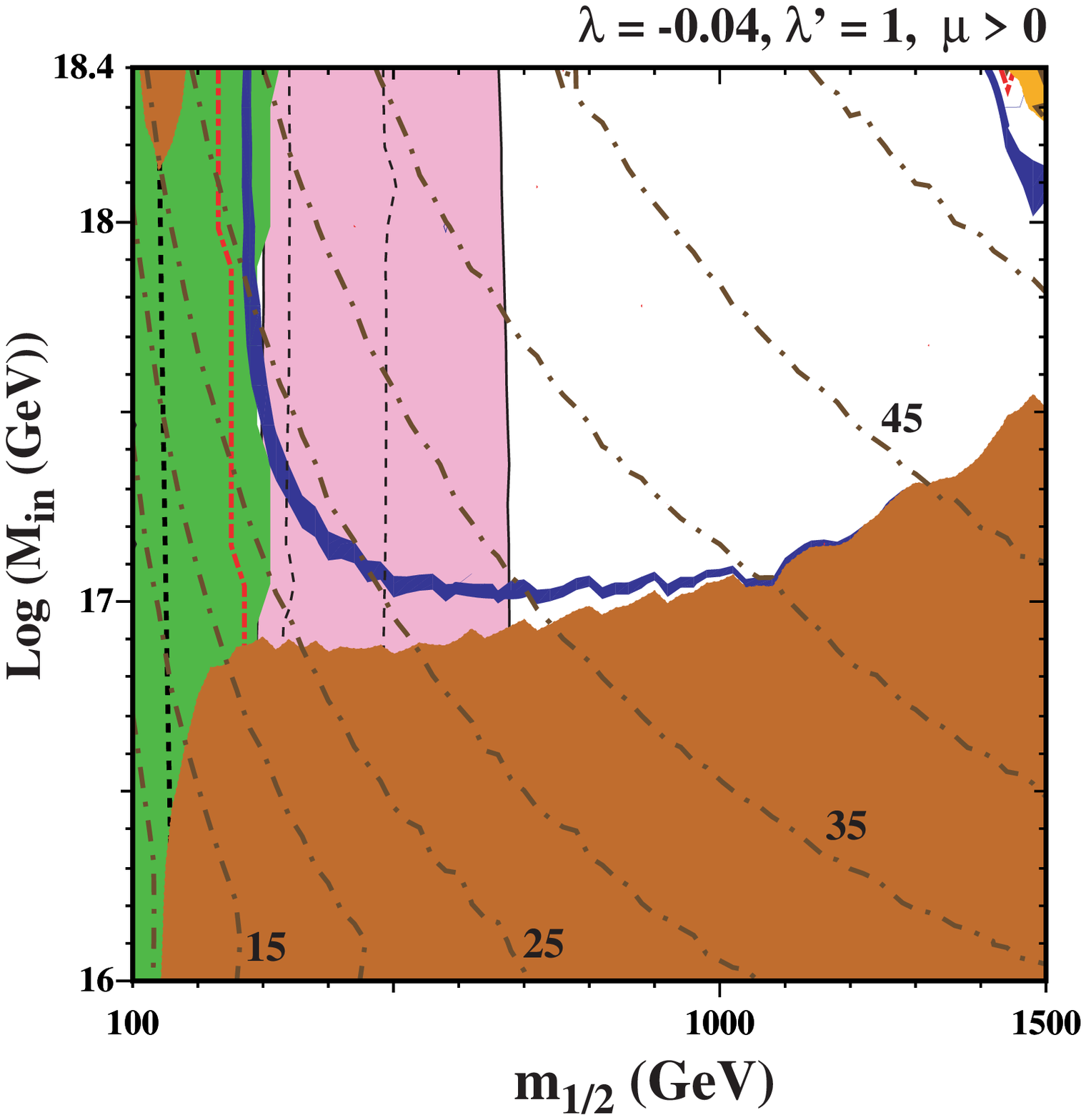,height=8cm}\\
\epsfig{file=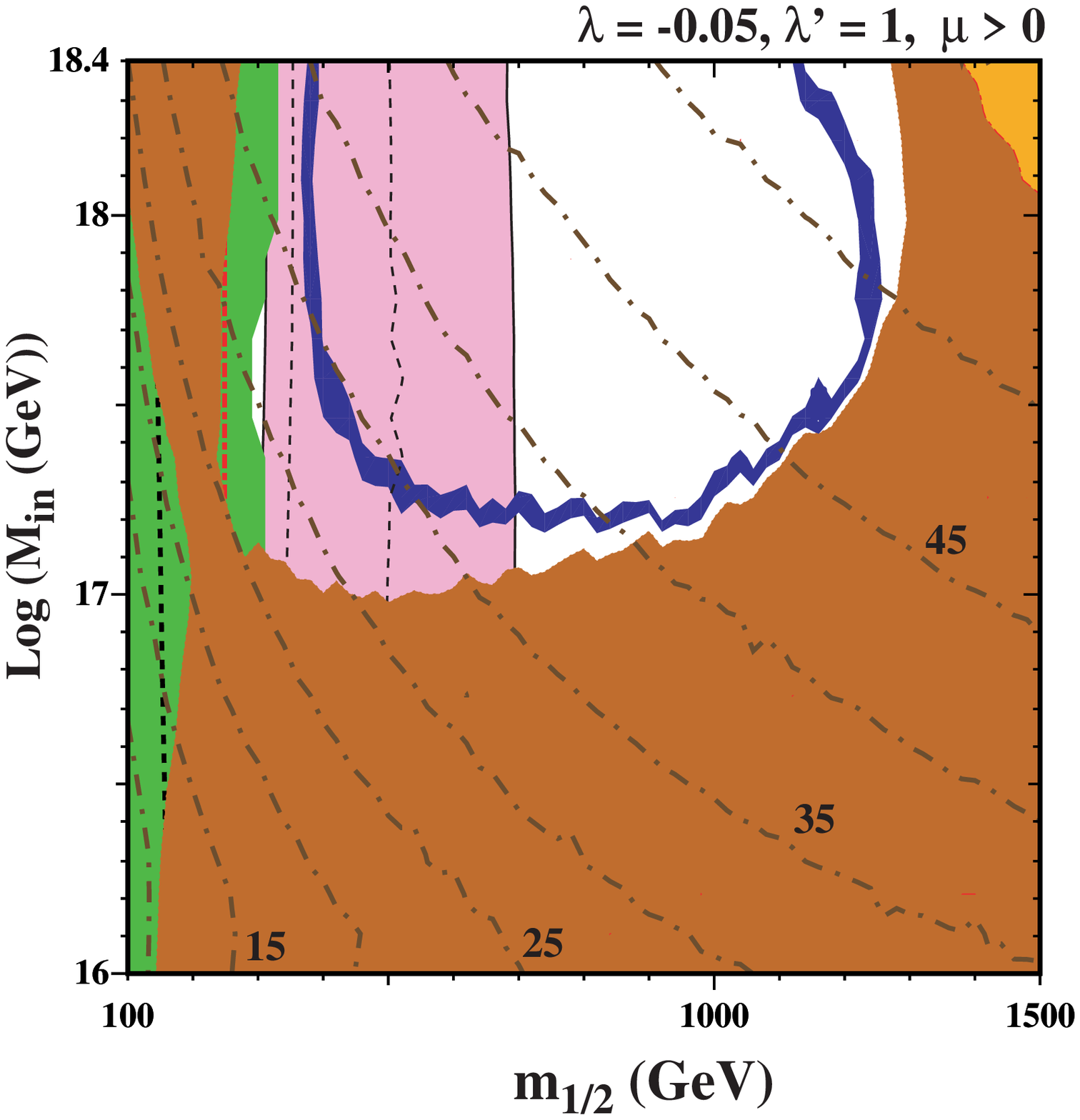,height=8cm}
\epsfig{file=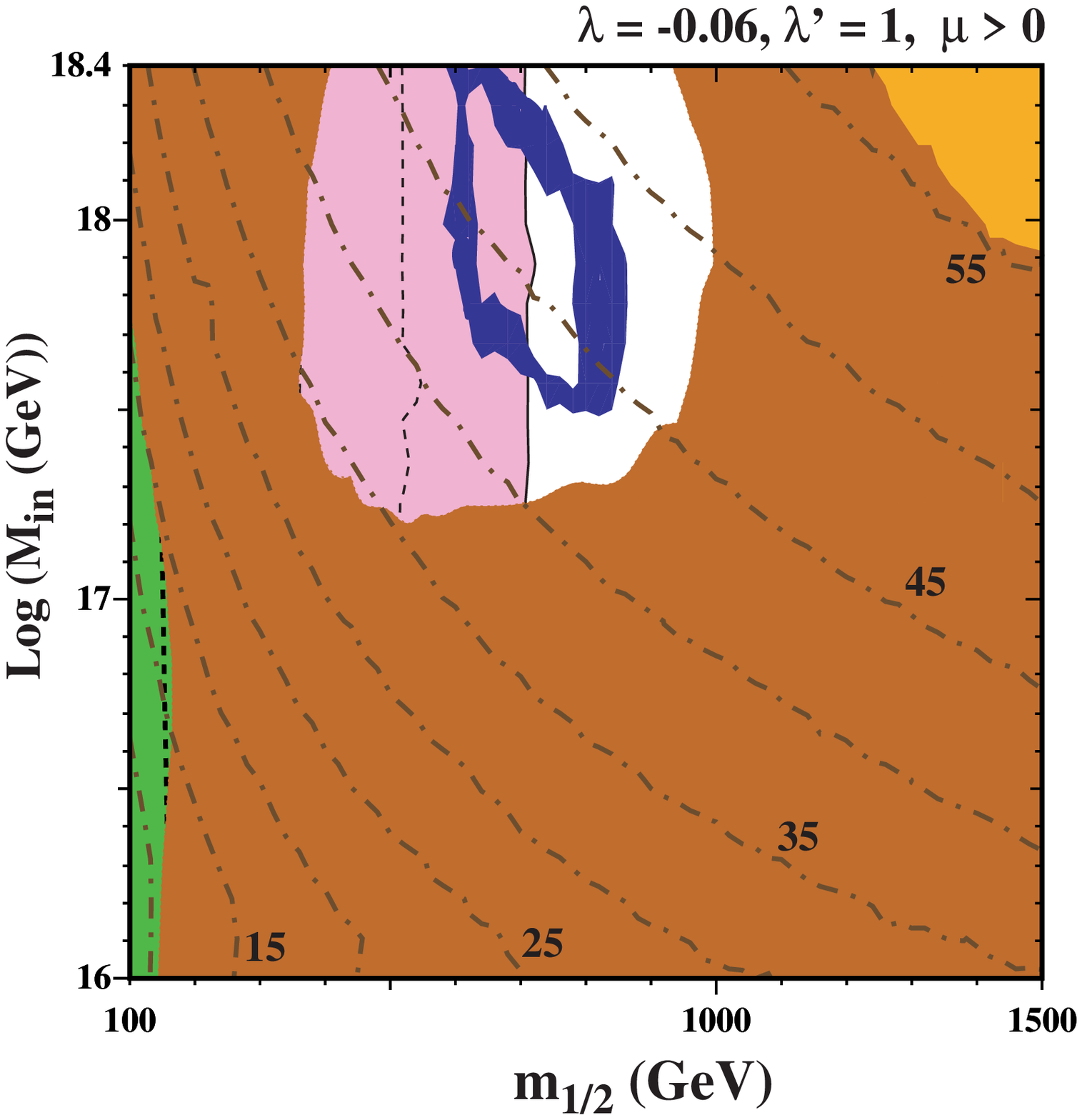,height=8cm}
\end{center}
\caption{\it
The $(m_{1/2}, M_{in})$ planes for the no-scale supergravity model with $\lambda' = 1$ and
(a) $\lambda = 0$, (b) $\lambda = -0.04$, (c) $\lambda = -0.05$ and (d)
$\lambda = -0.06$. The relic density $\ohsq$ is within the WMAP range in the blue strip. The
pink region between the black dashed (solid) lines is allowed by $g_\mu-2$ at 
the 1-$\sigma$ (2-$\sigma$) level. The diagonal dark brown dash-dotted lines are contours
of $\tan \beta$, as determined by the electroweak vacuum conditions.
The brown and green colored regions are excluded by the requirements of a neutralino LSP, and
by $b \to s \gamma$, respectively, and in the orange region we find no consistent solutions
to the RGEs. Areas to the left of the thick black dashed and red dash-dotted lines 
are ruled out by LEP searches for charginos and the lightest MSSM Higgs, respectively. More details can be found in the text.}
\label{fig:m12min1}
\end{figure}

\begin{figure}
\begin{center}
\epsfig{file=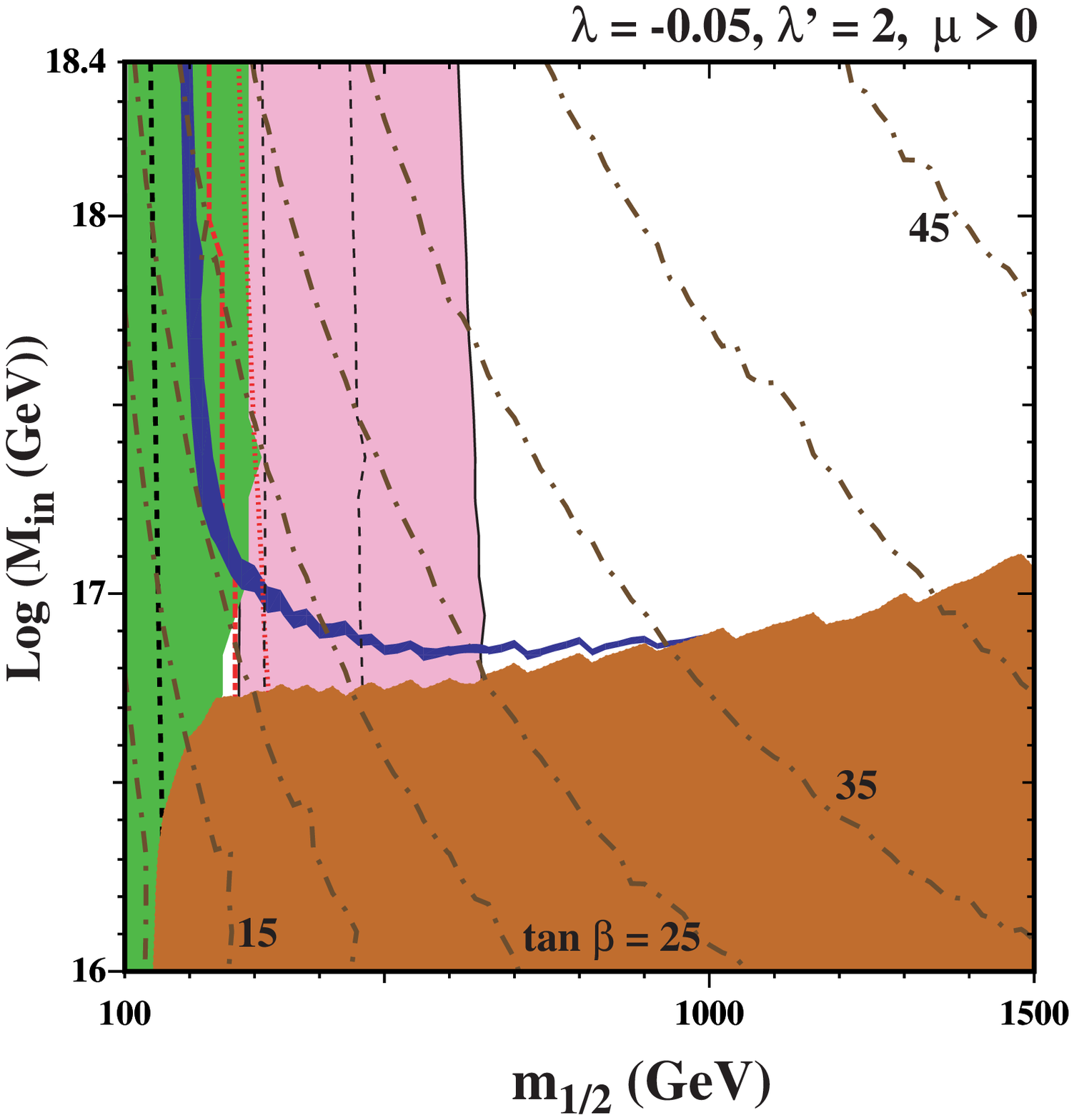,height=8cm}
\epsfig{file=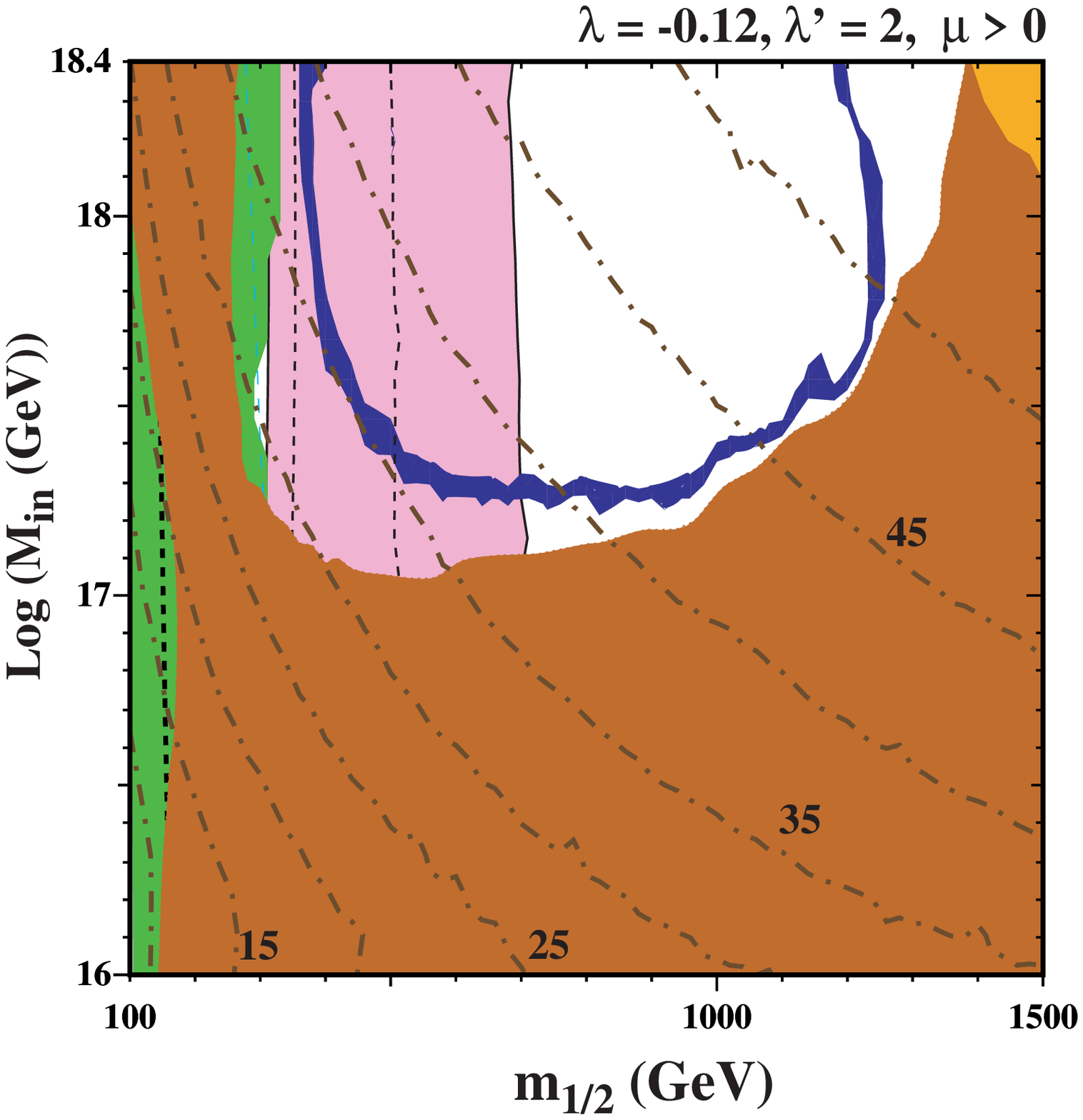,height=8cm}\\
\epsfig{file=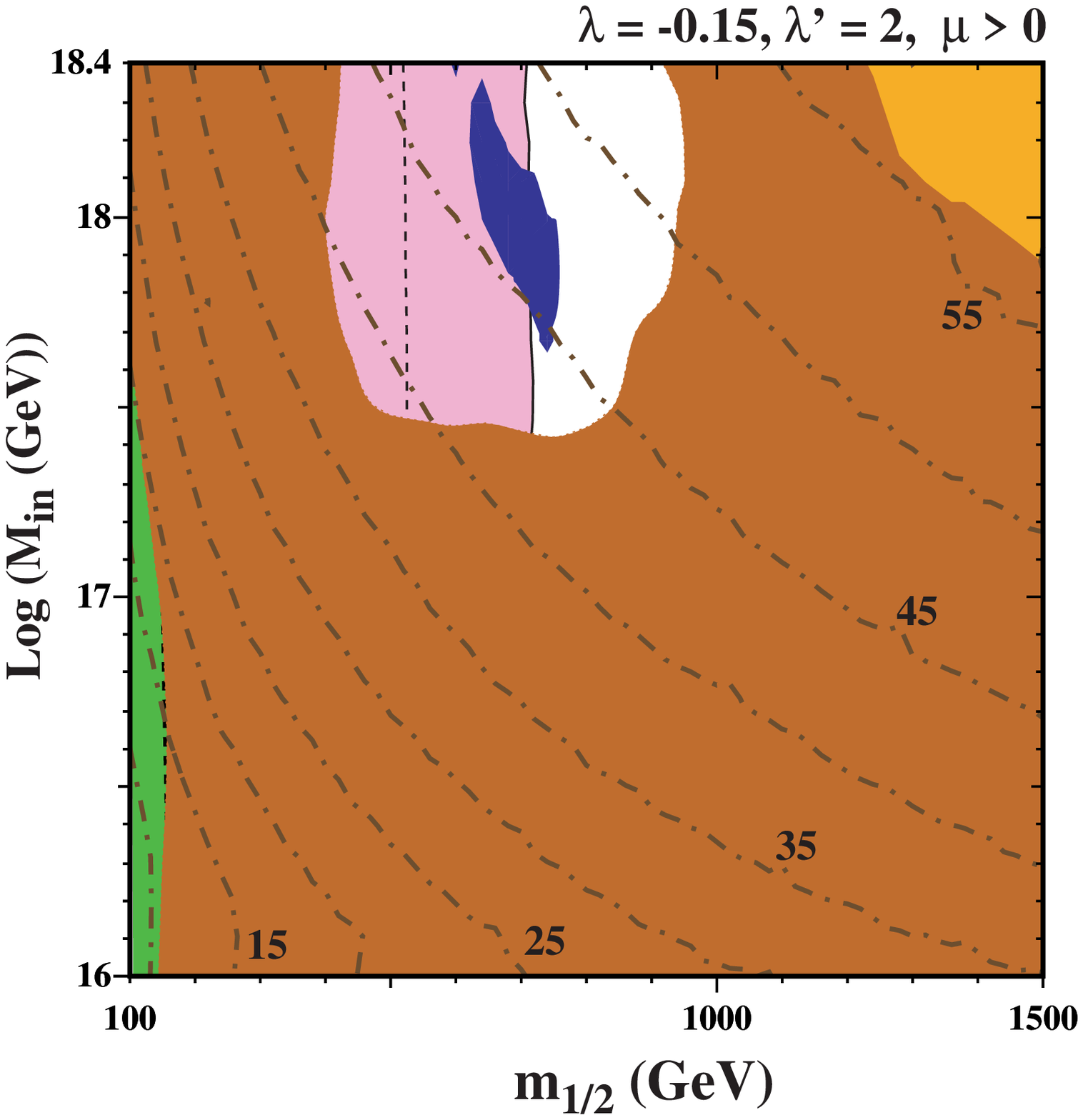,height=8cm}
\epsfig{file=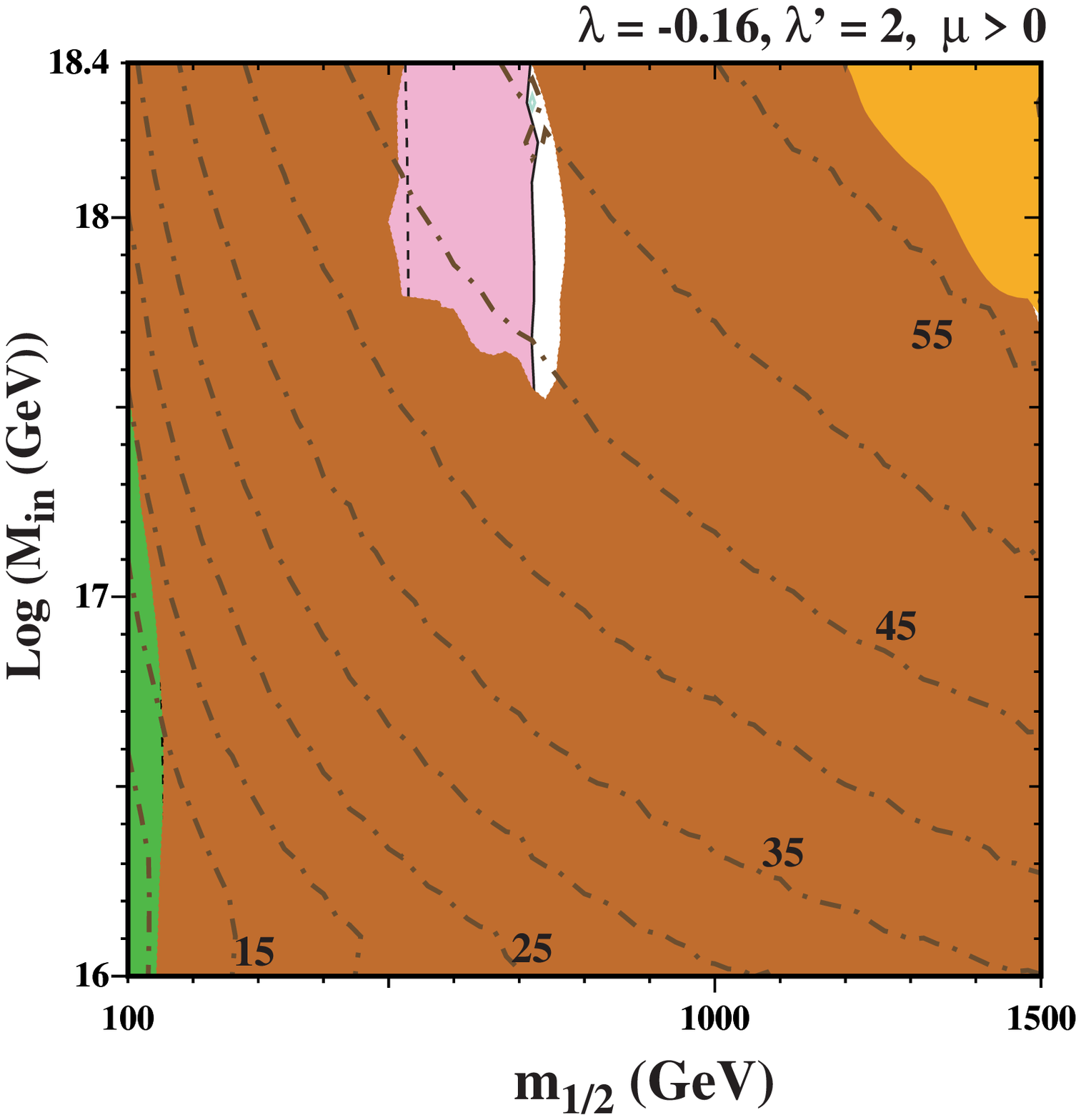,height=8cm}
\end{center}
\caption{\it 
Similar to Fig.~\protect\ref{fig:m12min1}, but for $\lambda' = 2$ and
(a) $\lambda = -0.05$, (b) $\lambda = -0.12$, (c) $\lambda = -0.15$ and (d) $\lambda = -0.16$.}
\label{fig:m12min2}
\end{figure}

We now discuss the 
$(m_{1/2}, M_{in})$ planes for the no-scale supergravity model with
various values of $\lambda$ and $\lambda'$ shown in Figs.~\ref{fig:m12min1} and \ref{fig:m12min2}, 
where we limit ourselves to $M_{in} \le \mplr$. These plots are for positive $\lambda'$ and negative $\lambda$: it is easy to check that the results are independent of the {\it overall} sign of $\lambda'$
and $\lambda$, but are sensitive to their {\it relative} sign. The plots in
Figs.~\ref{fig:m12min1} and \ref{fig:m12min2} are for the (more interesting) case
in which $\lambda'$ and $\lambda$ have opposite signs, we discuss later the (less
interesting) same-sign case. We recall that in all of the figures 
we have set $m_0 = A_0 = B_0 = 0$ and $\mu > 0$. We concentrate on values of
$\lambda' \ge 1$ since on the one hand, as we discuss below, 
the values of $\lambda$ allowed in no-scale models are typically $\ll \lambda'$ and, on
the other hand, small values of $\lambda$ are disfavoured by proton stability.
We are able to find stable
solutions of the RGEs for $\lambda' \lappeq 2.6$: the features seen in Figs.~\ref{fig:m12min1}
and \ref{fig:m12min2} persist up to the largest values of $\lambda'$ that we have studied.

For RGE calculations we used the program {\tt SSARD}~\cite{ssard}, 
which allows the computation of sparticle spectrum on the basis of
2-loop RGE evolution for the MSSM~\cite{Martin:1993zk} and 1-loop evolution
for minimal SU(5)~\cite{pp}, and perform cross-checks 
with {\tt ISAJET 7.80}~\cite{isajet} modified to include SU(5) running above $\mgut$.
We set soft parameters at $M_{in}$ according to Eqs.(\ref{BC1}) and perform matching 
between SU(5) and MSSM according to expressions (\ref{matching1},\ref{Bmatch}) at the scale $\mgut$.
The location of the latter is determined dynamically as the scale where $g_1=g_2$,
and is approximately
$1.5 \times 10^{16}$~GeV. 
Throughout the paper we assume $m_t = 173.1$~GeV~\cite{mt} and
$m_b^{\overline{MS}}(m_b) = 4.2$~GeV~\cite{rpp}. 

In each plane of 
Figs.~\ref{fig:m12min1} and \ref{fig:m12min2},
we indicate by brown shading the region that is excluded because the LSP is the lighter stau  
$\stau_1$, and the region where we find no consistent solution of the RGEs
is indicated by orange shading. Our treatment of
$BR(b \to s \gamma)$ follows that in~\cite{gam,bsgprocedure}, and the
region excluded at the 95\% CL~\cite{hfag} is shaded green.
The region favoured by $g_\mu - 2$ measurements~\cite{newBNL,g-2} if the Standard Model contribution is calculated using low-energy $e^+ e^-$ data~\cite{Davier} is shaded pink, with the
$\pm 1\mhyphen\sigma$ contours shown as black dashed lines and the $\pm 2\mhyphen\sigma$ contours shown as solid black lines.
The LEP lower limit on the chargino mass~\cite{LEPsusy} is shown as a thick black dashed
line, and the experimental lower limit on $m_h$~\cite{LEPHiggs} is indicated as a red 
dash-dotted line. This
shows the position of the 95\% confidence-level lower limit obtained by
combining the experimental likelihood from direct searches at LEP~2 and a global
electroweak fit, convolved with the theoretical and parametric errors in $m_h$~\footnote{
We thank A.~Read for providing the LEP $CL_s$ values.}, which provides a more exact
(and relaxed) interpretation of the nominal LEP Higgs limit of 114.4~GeV within the MSSM.
In the models under study, the
couplings of the lightest supersymmetric Higgs are very similar to those in the Standard
Model, so the same nominal lower limit $m_h > 114.4$~GeV applies. However, the light Higgs 
mass is computed in terms of the model parameters using the {\tt FeynHiggs~2.6.5} 
code~\cite{FeynHiggs}, whose nominal results are assigned a theoretical error $\sim 1.5$~GeV
in drawing the exclusion contour, moving it towards smaller $m_{1/2}$
by $\sim 50$~GeV~\footnote{The
constraint that would be obtained with the nominal value $m_h = 114.4$~GeV is also shown, for
comparison, as a dotted red line in the
top left panels of Figs.~\ref{fig:m12min1} and \ref{fig:m12min2}.}. 
Finally, we use blue colour to indicate the regions where the
neutralino relic density falls within the 2-$\sigma$ WMAP range~\cite{WMAP}, 
$0.097 \leq \Omega_{CDM}h^2 \leq 0.122$.

The diagonal dark brown dash-dotted lines in Figs.~\ref{fig:m12min1} and \ref{fig:m12min2}
are contours of $\tan \beta$, rising in steps of 5 from left to right. The
intercepts of the $\tan \beta$ contours on the $m_{1/2}$ axis are the same in all panels, 
{\it e.g.}, $\tan \beta = 15$ when
$m_{1/2} \simeq 260$~GeV and $\tan \beta = 25$ when $m_{1/2} \simeq 700$~GeV, but the
slopes of the contours depend on the no-scale parameters $\lambda, \lambda'$, rotating
counter-clockwise as $\lambda$ decreases for any fixed value of $\lambda'$.
This behaviour arises because the value of $\tan \beta$ for any point on the plane is 
sensitively tied to the matching 
condition in Eq.~(\ref{Bmatch}).  In Ref.~\cite{vcmssm}, it was shown that GUT-scale value of the
MSSM $B$ parameter is almost always a monotonically-increasing function of $\tan \beta$ in the
CMSSM.  While its slope with respect to $\tan \beta$ is steep at low $\tan \beta$,
it increases very little for $\tan \beta \gappeq 10$. As a result, relatively small
changes in the required value for $B_{\mgut}$ could entail very large changes in $\tan \beta$. 

From the RGEs of $\alamp$, $\mSig^2$, $B_H$ and $B_\Sigma$ in Eqs.~(\ref{RGEb}, \ref{morerge})
we can understand the qualitative behaviour of the $\tan \beta$ contours.
Since our boundary conditions at $M_{in}$ correspond to $A_0 = 0$ and $\mSig^2 = 0$,
the initial driving force in all of the relevant RGEs comes from the non-zero gaugino mass, $M_5$,
driving $\alamp$, $B_H$, and $B_\Sigma$ positive and driving the dominant term
$\mSig^2$ positive. As $M_{in}$ is increased, the right-hand side of the matching condition (\ref{match}) 
is driven higher (for $\lambda/\lambda' < 0$), resulting in a larger value for $\tan \beta$.  Similarly,
a larger value for $m_{1/2}$ ($M_5$) leads to a larger initial kick in the RGE evolution,
which also increases the right-hand side of Eq.~(\ref{match}) and hence leads to higher $\tan \beta$. 
This is precisely the pattern we see in Figures \ref{fig:m12min1} and \ref{fig:m12min2}.
At very large $M_{in}$ and/or $m_{1/2}$, $\tan \beta$ is pushed above 55, and 
soon thereafter we are not able to obtain solutions to the MSSM RGEs.  This area 
is shaded orange.  Furthermore, we see that as we increase $-\lambda$ (for fixed
$\lambda'$) we again increase the right-hand side of Eq.~(\ref{match}) and 
the $\tan \beta$ contours rotate counter-clockwise, so that we obtain higher $\tan \beta$
for any given choice of ($m_{1/2}$, $M_{in}$).

The $(m_{1/2}, M_{in})$ planes in Fig.~\ref{fig:m12min1} are for $\lambda' = 1$ and (a)
$\lambda = 0$, (b) $\lambda = -0.04$, (c) $\lambda = -0.05$ and (d) $\lambda = -0.06$.
We see that in the first two panels of Fig.~\ref{fig:m12min1} the WMAP-compatible region
takes the form of a thin L-shaped strip in the $(m_{1/2}, M_{in})$ plane with a rounded corner:
points above and to the right of the L have values of $\ohsq$ that are too large. In panel (a)
for $\lambda = 0$,
the near-horizontal part of the line is located at $M_{in} \sim 5 \times 10^{16}$~GeV and 
extends from $m_{1/2} \sim 200$~GeV to $\sim 1000$~GeV, larger values being
excluded by the requirement that the LSP not be charged, and 
we find that $\tan \beta \in (16, 30)$. 
All the base strip is compatible with the LEP chargino
constraint, and with $b \to s \gamma$. However, only the portion with $m_{1/2} \gappeq
300$~GeV is compatible with the LEP lower limit on $m_h$, taking into
account the theoretical uncertainty in the {\tt FeynHiggs} calculation of $m_h$, and
the near-vertical part of the no-scale strip is always
incompatible with the LEP Higgs constraint and (mostly) $b \to s \gamma$. 

In panel (b) of Fig.~\ref{fig:m12min1} for $\lambda = -0.04$,
the near-vertical part of the WMAP-compatible strip has moved to larger
values of $m_{1/2} \sim 300$~GeV, and is now compatible 
with the LEP Higgs constraint and (mostly) the $b \to s \gamma$ constraint. Also,
the values of $M_{in}$ along the near-horizontal part of the allowed
strip rise to $M_{in} \gappeq  10^{17}$~GeV, values of 
$m_{1/2} \lappeq 1300$~GeV have a neutralino LSP, and $\tan \beta \lappeq 45$.
We also note in the upper right corner of panel (b) of Fig.~\ref{fig:m12min1} the
appearance of a portion of another WMAP-compatible strip at $\tan \beta > 50$.

Panel (c) of Fig.~\ref{fig:m12min1} for $\lambda = -0.05$ has a rather 
different appearance, but is in fact
a natural continuation of the trends seen in the previous panels. In particular, the 
near-vertical part of the WMAP-compatible strip has moved to
larger $m_{1/2} \sim 400$~GeV, and is compatible with both $m_h$ and
$b \to s \gamma$, and the near-horizontal part of the strip has risen to
$M_{in} \sim 2 \times 10^{17}$~GeV.
More dramatically, the WMAP-compatible strip now becomes a (rounded)
triangle, with a hypotenuse connecting the previous two strips at relatively
large $m_{1/2}$ and $M_{in}$ and $\tan \beta \gappeq 50$
(though the triangle closes only when $M_{in} > \mplr$),
expanding the fragment of the 
hypotenuse visible in the upper right corner of panel (b) of of Fig.~\ref{fig:m12min1}.
We emphasize that all the 
triangle has a neutralino LSP, and that the stau-LSP contour loops around the
base and hypotenuse: the region within the triangle has too much dark matter.

Turning finally to panel (d) of Fig.~\ref{fig:m12min1} for $\lambda = -0.06$, we see
that the triangle has contracted to an `doughnut' with $m_{1/2} \in (600, 900)$~GeV,
$M_{in} \in (3 \times 10^{17}, 3 \times 10^{18})$~GeV and $\tan \beta \in (42, 49)$, 
all of which is again compatible with the
LEP Higgs constraint and $b \to s \gamma$, and sits partly in the 
region preferred by $g_\mu - 2$. Once more, the stau-LSP contour loops 
around the doughnut. 
As $\lambda$ is dialed to more negative values the doughnut continues to shrink and eventually disappears for $\lambda \lappeq -0.07$.

A similar evolutionary pattern as $\lambda$ decreases is seen in Fig.~\ref{fig:m12min2},
where we fix $\lambda' = 2$. In panel (a) for $\lambda = -0.05$, the WMAP-compatible
strip is again L-shaped, with a near-vertical part at $m_{1/2} \sim 200$~GeV
that is incompatible with the Higgs constraint and $b \to s \gamma$,
and a near-horizontal part at $M_{in} \sim 8 \times 10^{16}$~GeV that
extends to $m_{1/2} \sim 1000$~GeV before the LSP ceases to be a neutralino.
At  $\lambda = -0.12$ the region allowed by the relic density has the same (rounded) 
triangular form as panel (c) of Fig.~\ref{fig:m12min1} (the triangle again closes only 
when $M_{in} > \mplr$). We see again the feature of the
stau-LSP contour looping around the WMAP-compatible strip. In panel (c)
for $\lambda = -0.15$ we see just a `blob' with $m_{1/2} \in (500, 650)$~GeV
and $M_{in} \in (5 \times 10^{17}, 3 \times 10^{18})$~GeV on the edge of the region 
preferred by $g_\mu - 2$. This remaining `blob' 
disappears for larger $-\lambda$ as shown in panel (d) for  $\lambda = -0.16$.
Here, the area within the `window' is phenomenologically allowed, though the
relic density lies below the WMAP range.

\begin{figure}
\begin{center}
\epsfig{file=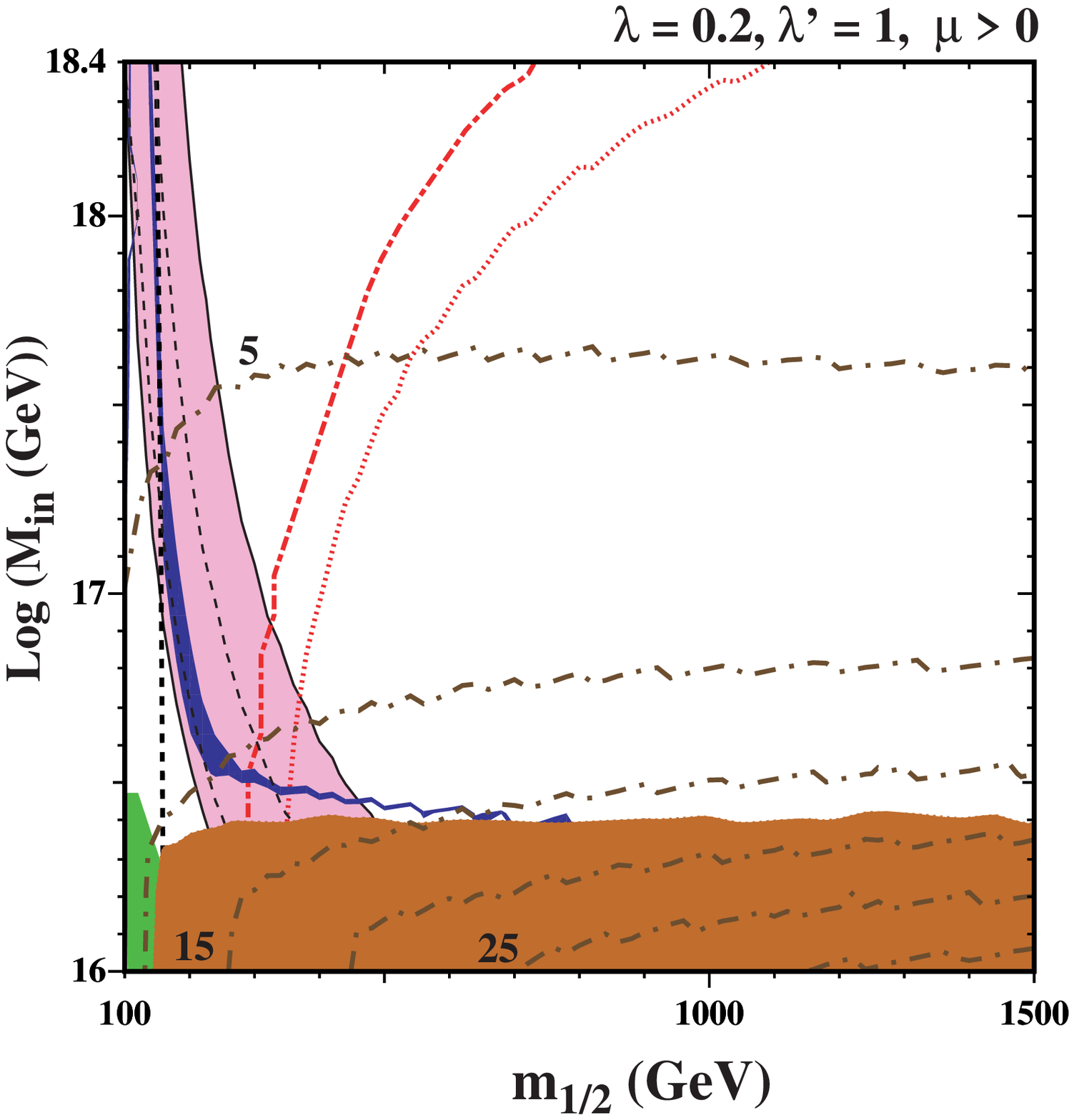,height=7.5cm}
\epsfig{file=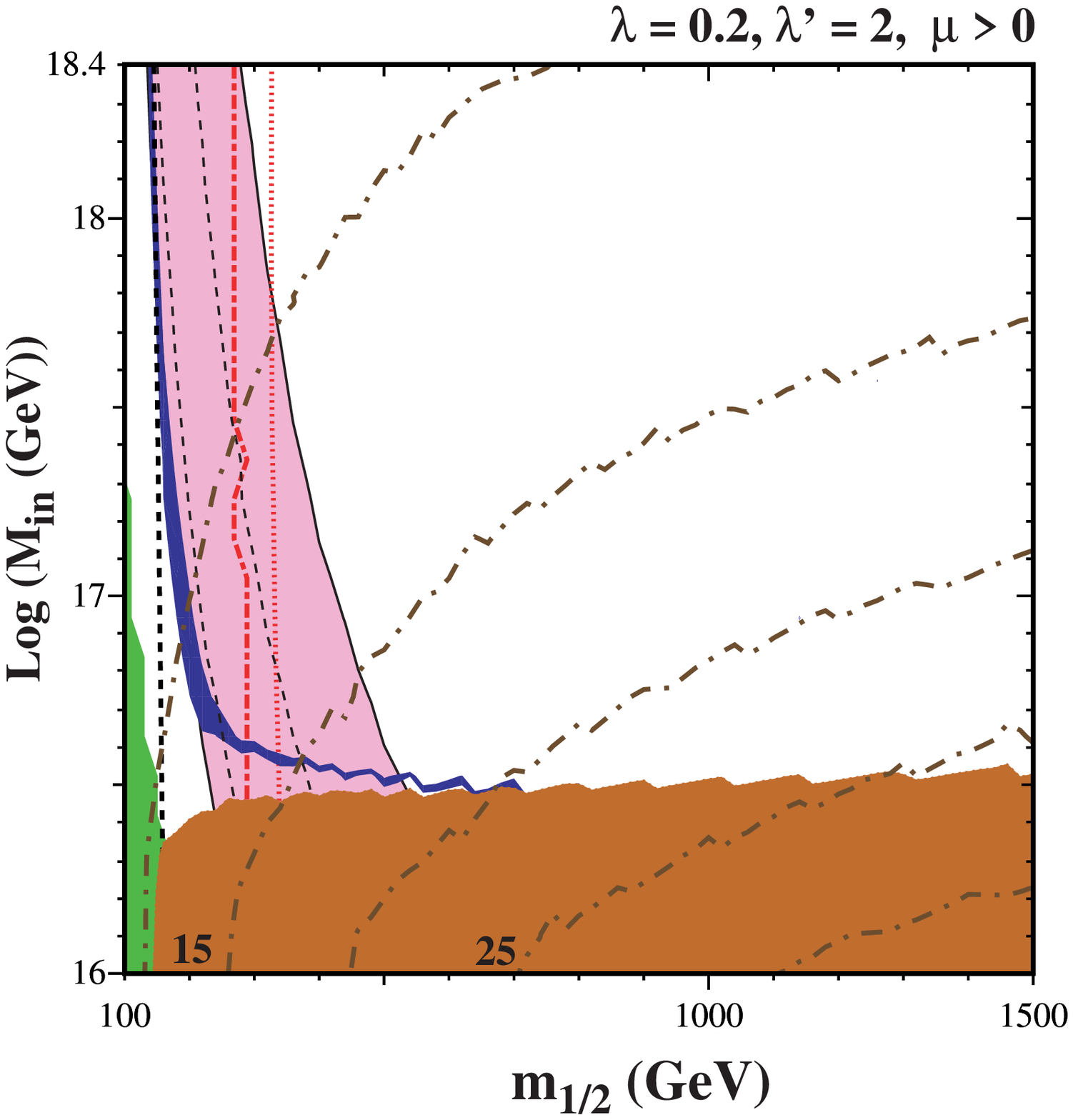,height=7.5cm}\\
\end{center}
\caption{\it
The $(m_{1/2}, M_{in})$ planes for $(\lambda, \lambda') = (0.2, 1)$ and $(0.2, 2)$ in the left and right plots, respectively. 
The curves and shadings have the same interpretations as for Fig.~\protect\ref{fig:m12min1}.}
\label{fig:positive}
\end{figure}

So far, we have displayed the regions of no-scale parameter space allowed when
$\lambda$ and $\lambda'$ have opposite signs. We now discuss the same-sign case, 
based on the examples shown in
Fig.~\ref{fig:positive}. As $\lambda/\lambda'$ increases
and becomes positive, the matching conditions (\ref{match}, \ref{Bmatch})
drive the slopes of the $\tan \beta$ contours to decreasing positive values.
That is, as $\lambda/\lambda' > 0 $ increases, the right-hand side of eq. (\ref{Bmatch})
decreases, and the $\tan \beta$ contours rotate clockwise, leading to
smaller values of $\tan \beta$ for fixed ($m_{1/2}, M_{in}$).
The region of parameter space allowed by the dark matter density constraint
retains its L shape in this limit, with the base moving to lower $M_{in}$ and the
near-vertical arm to smaller $m_{1/2}$, as seen for the examples
$(\lambda, \lambda') = (0.2, 1)$ and $(0.2, 2)$ shown in the left and right plots of
Fig.~\ref{fig:positive}, respectively. Because the vertical strip is situated at relatively
low $\tan \beta$, it is excluded by the LEP Higgs mass constraint in both examples shown.

\begin{figure}
\begin{center}
\epsfig{file=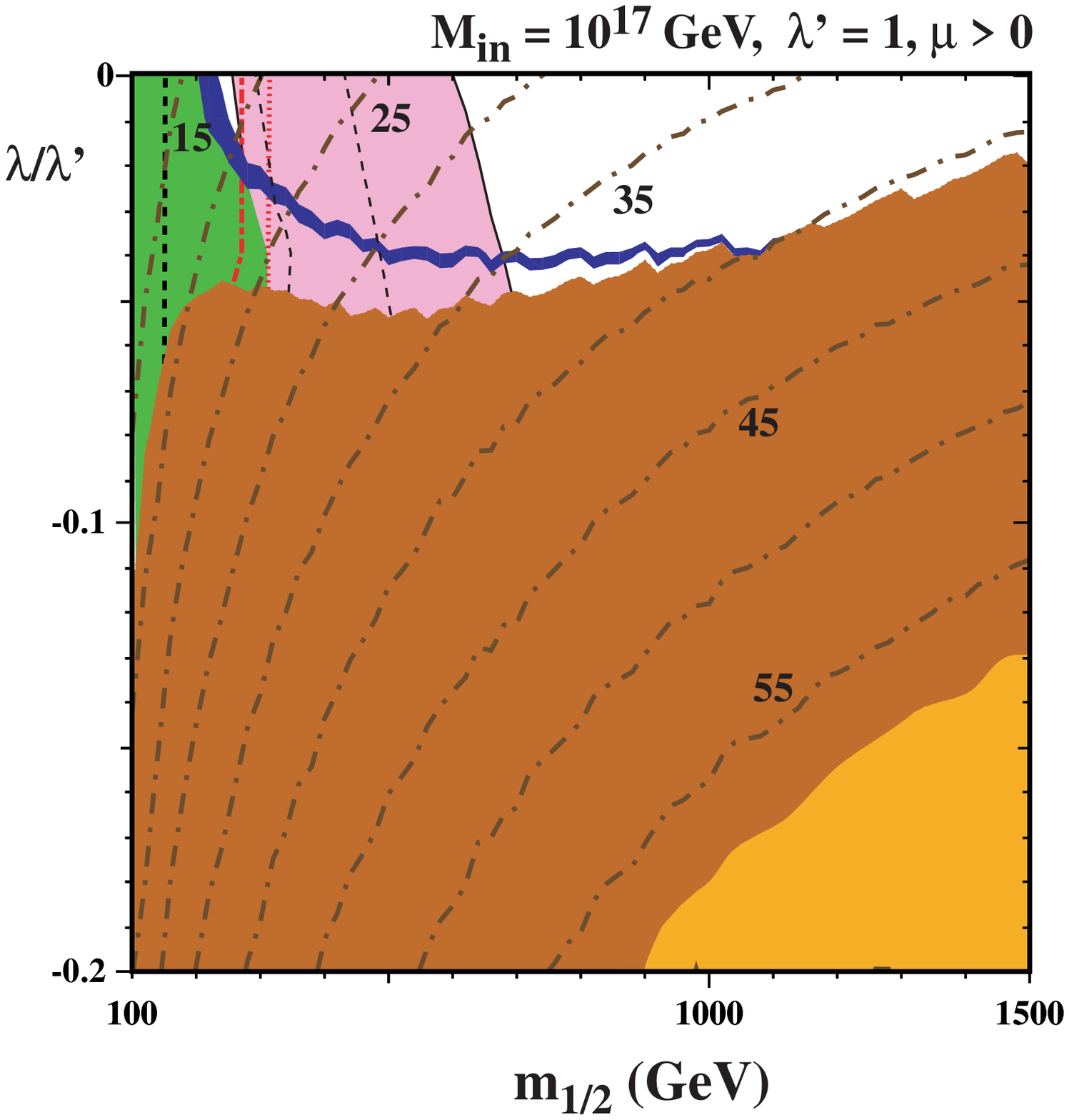,height=8cm}
\epsfig{file=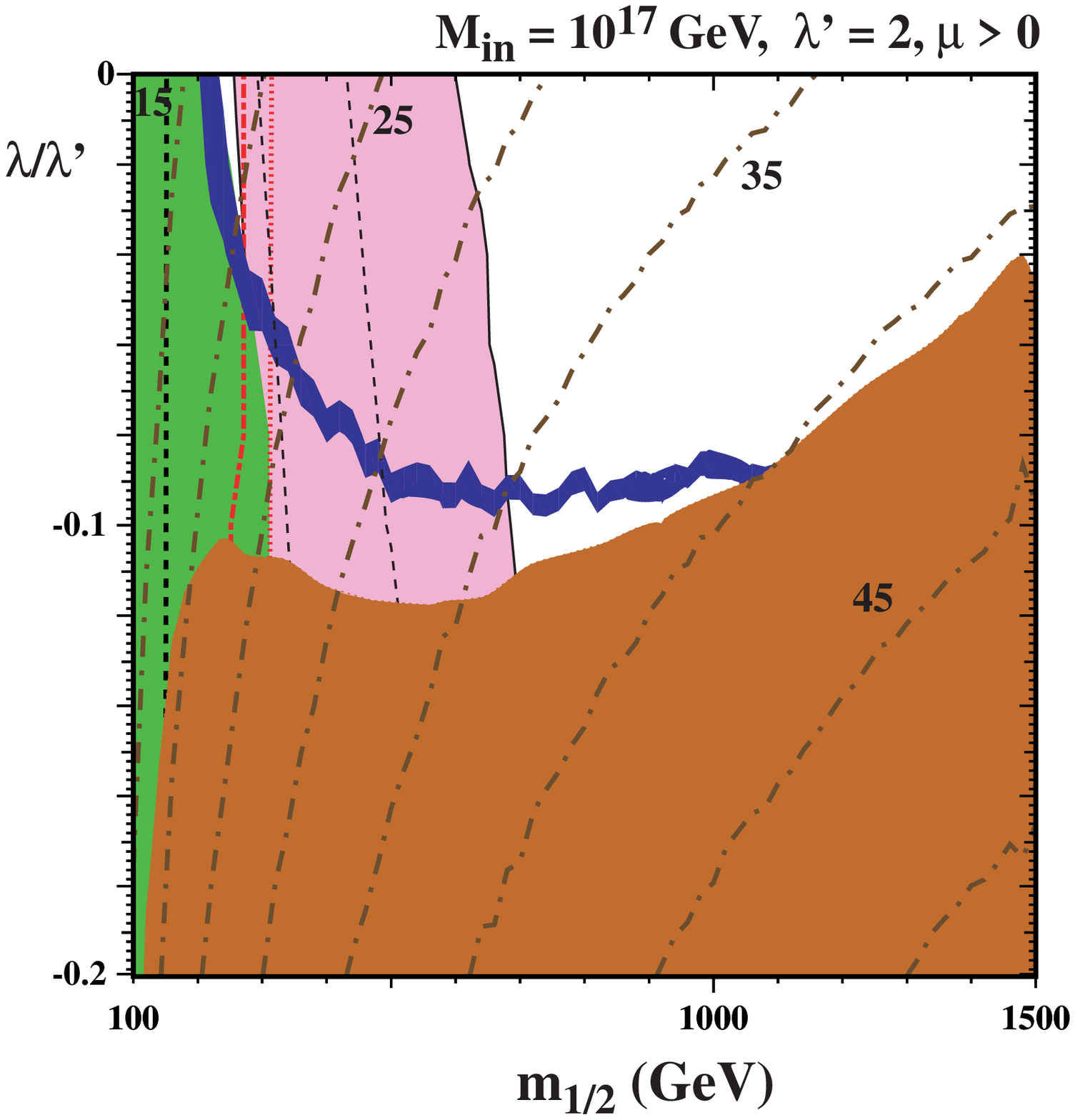,height=8cm}\\
\epsfig{file=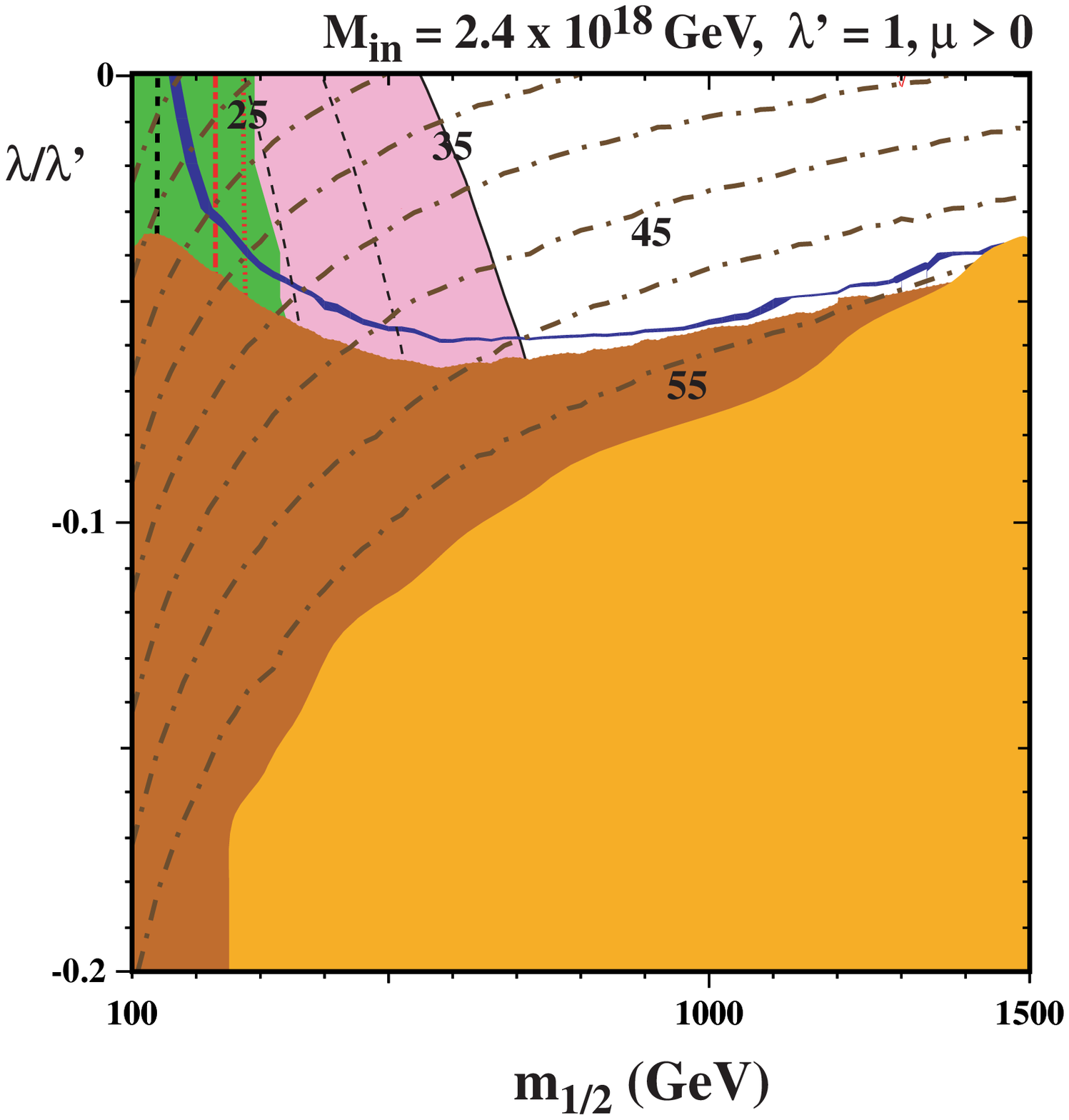,height=8cm}
\epsfig{file=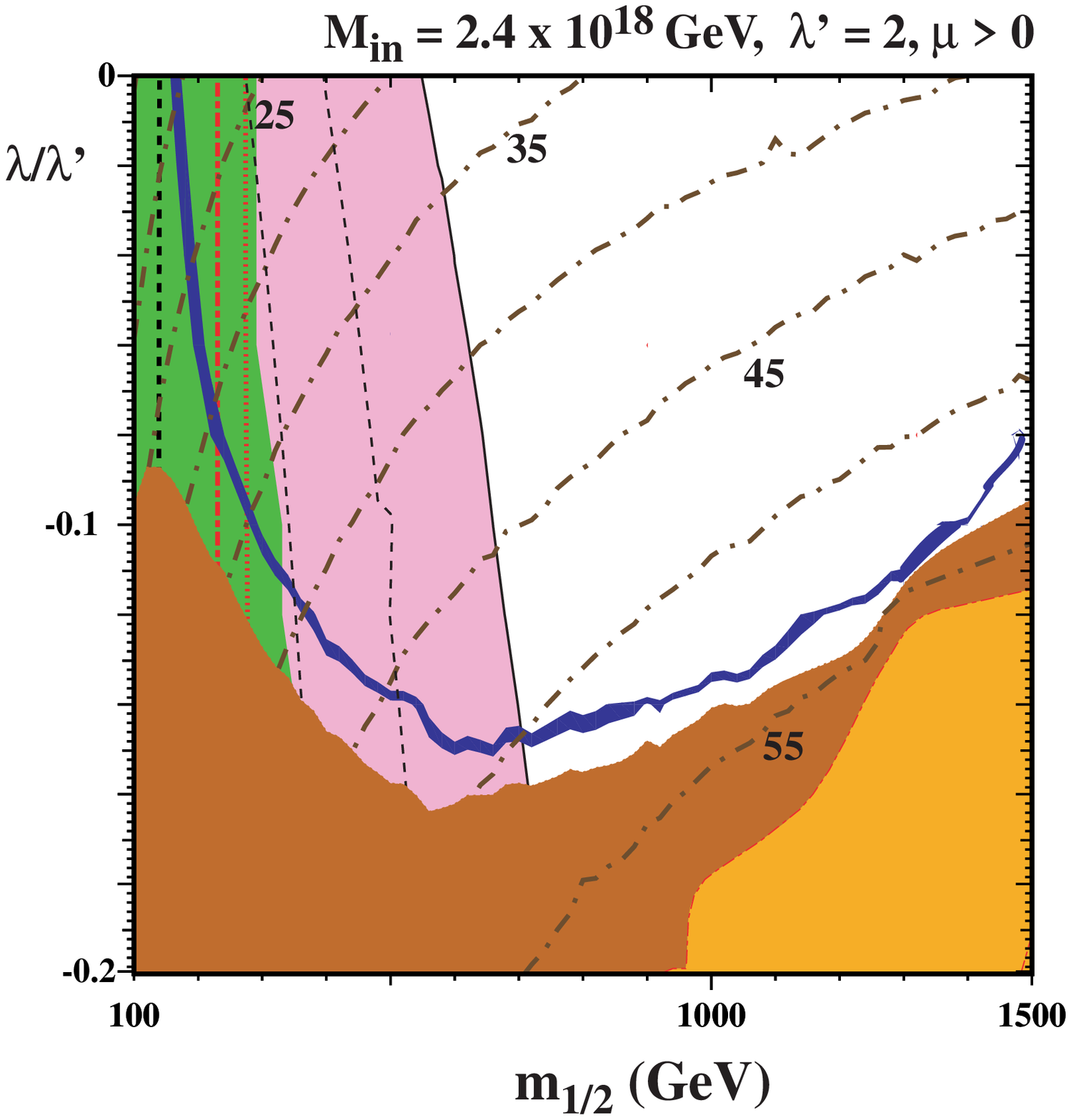,height=7.8cm}
\end{center}
\caption{\it
The $(m_{1/2}, \lambda)$ planes for $M_{in} = 10^{17}$~GeV (upper plots) and $M_{in} = \mplr$
(lower plots), for $\lambda' = 1$ (left plots) and $\lambda' = 2$ (right plots). 
The curves and shadings have the same interpretations as for Fig.~\protect\ref{fig:m12min1}.}
\label{fig:m12minlrat}
\end{figure}

Fig.~\ref{fig:m12minlrat} displays the extensions of these results to the $(m_{1/2}, \lambda)$
planes for the fixed values $M_{in} = 10^{17}$~GeV (upper plots) and $M_{in} = \mplr$
(lower plots), for $\lambda' = 1$ (left plots) and $\lambda' = 2$ (right plots). We see that for $M_{in}
= 10^{17}$~GeV and $\lambda' = 1$ the WMAP-compatible strip has $\lambda \sim -0.04$,
whereas values of $-\lambda$ about twice as large are required if $\lambda' = 2$.
A similar change in $\lambda$ is seen in the lower plots for $M_{in} = \mplr$. On the
other hand, the required values of $\lambda$ do not change greatly between the
two values of $M_{in}$. These plots therefore confirm the earlier finding that
consistent no-scale models require $\lambda \ll \lambda'$, whatever the assumed
values of $\lambda'$ and $M_{in} \gg M_{GUT}$, and that the results are more sensitive to
$\lambda$ than to $\lambda'$.  If extended to positive values of $\lambda/\lambda'$ for these values
of $M_{in}$, these plots would show a continuation of the vertical relic 
density strips at low $m_{1/2}$ which are
excluded by the Higgs mass constraint.  As $M_{in}$ is lowered further (below $10^{17}$ GeV),
the horizontal part of the L moves towards and into the positive $\lambda/\lambda'$ domain.

\section{Sparticle Spectra in No-Scale Models}

\begin{figure}[ht]
\begin{center}
\epsfig{file=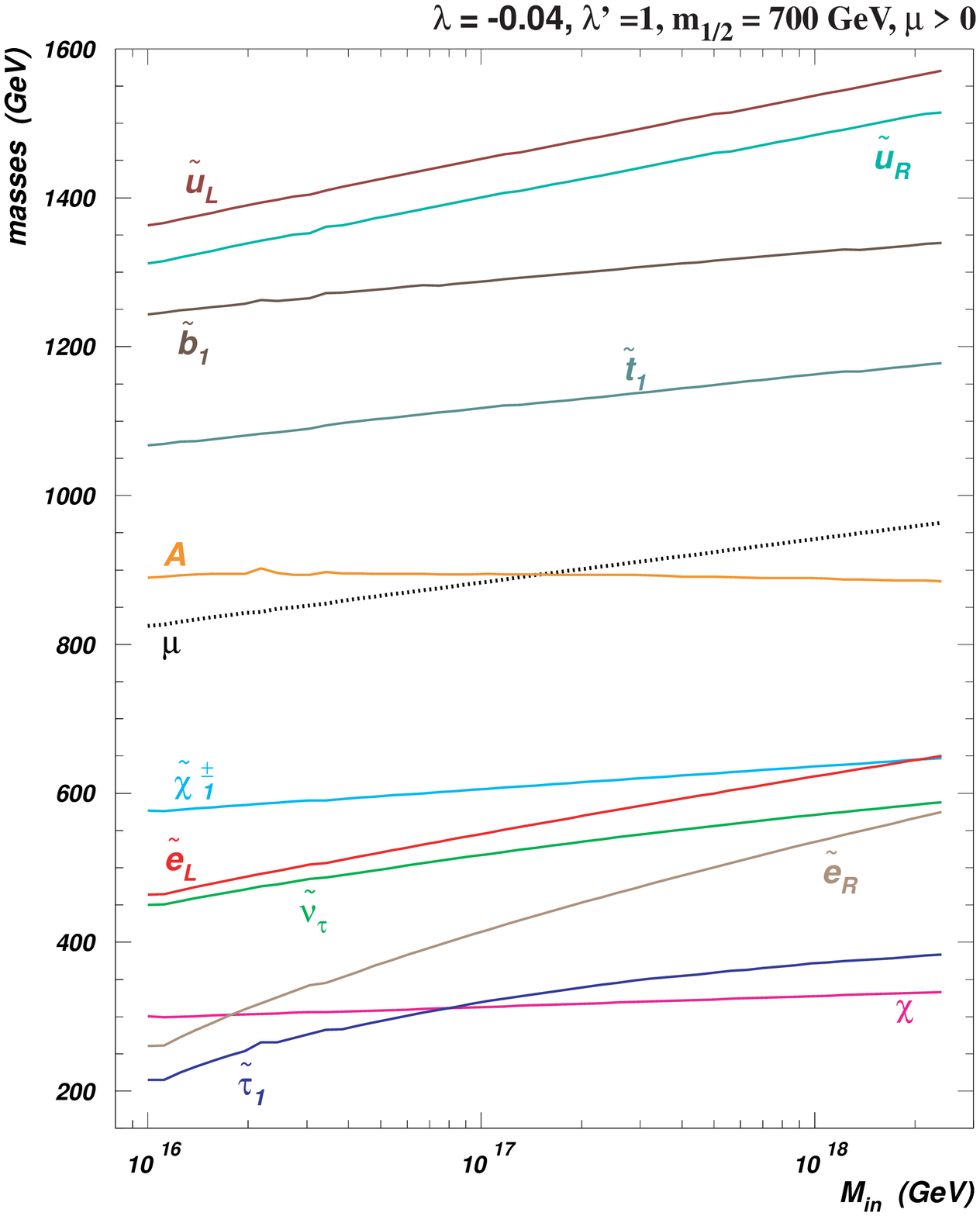,height=8.5cm}
\hskip .1in
\epsfig{file=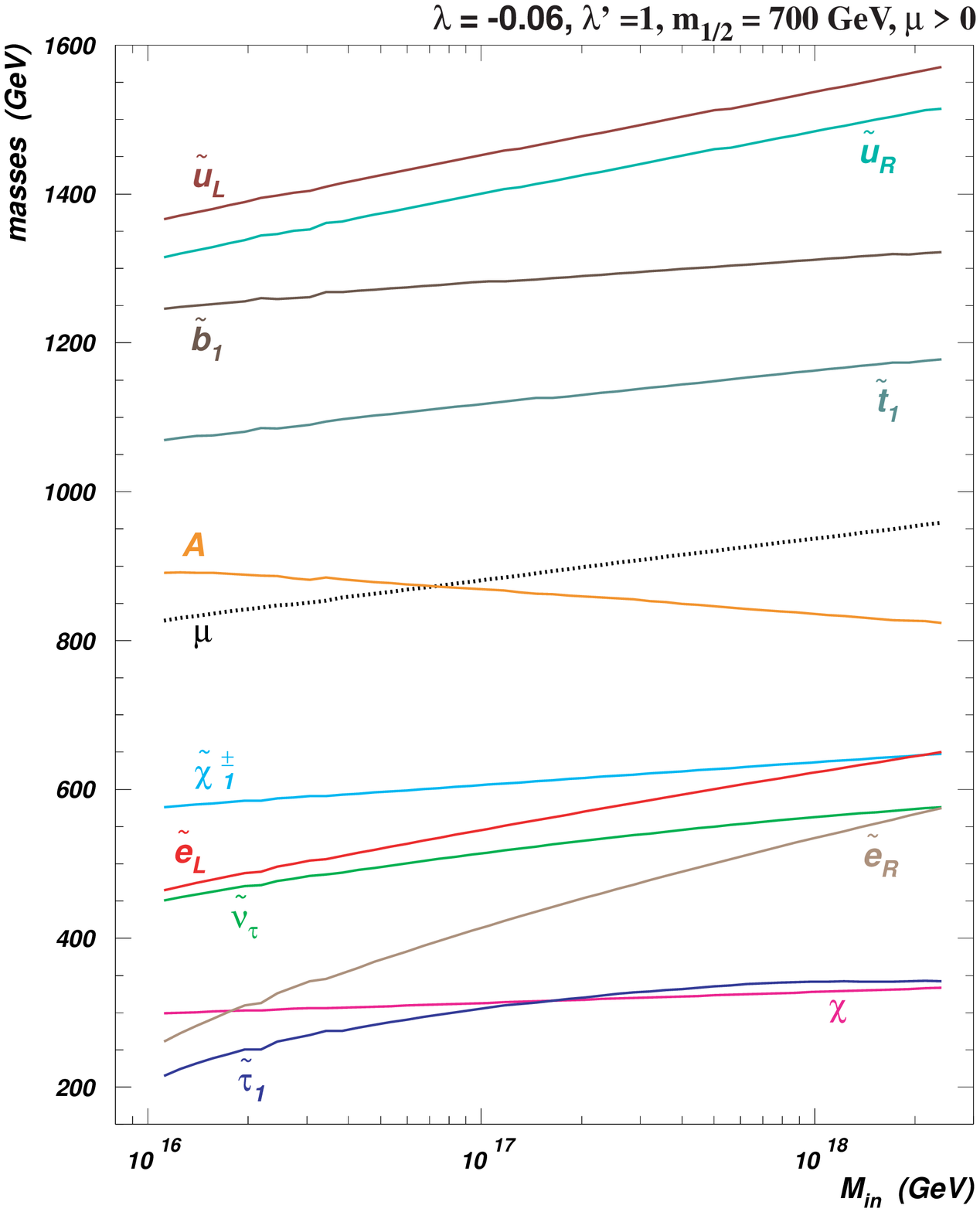,height=8.5cm}
\end{center}
\caption{\it
The spectra in the no-scale SU(5) model for $\lambda' = 1$ and $m_{1/2} = 700$~GeV
and (left) $\lambda = -0.04$ and (right) $\lambda = -0.06$, as functions of $M_{in}$. Note that
in each case there is an unacceptable
range of low $M_{in}$ where $m_{\tilde \tau_1} < m_\chi$, followed by
range of higher $M_{in}$ where $m_{\tilde \tau_1} > m_\chi$. 
There are viable
coannihilation strips in the intermediate range of $M_{in}$ near the crossing points.}
\label{fig:staumass}
\end{figure}

Many of the features in Figs.~\ref{fig:m12min1} and \ref{fig:m12min2} can be
understood by looking at the sparticle spectra. 
In Fig.~\ref{fig:staumass}, we show how it changes with $M_{in}$ 
for the specific choices $m_{1/2} = 700$~GeV,
$\lambda' = 1$ and (left) $\lambda = -0.04$, (right) $\lambda = -0.06$~\footnote{Analogous
plots for other values of $m_{1/2}$ and $\lambda'$ show similar features.}. We see in the
left panel that the lighter stau, $\stau_1$, is the LSP and is lighter than the lightest neutralino, $\chi$, for $M_{in} \lappeq 8 \times 10^{16}$~GeV, but increases monotonically in mass
while $m_\chi$ remains roughly constant as $M_{in}$ increases. The near-horizontal
part of the WMAP-compatible strip in panel (b) of Fig~\ref{fig:m12min1} appears due to 
stau-$\chi$ coannihilation~\cite{stauco}, which is important when the stau is only slightly heavier
than the $\chi$, for $M_{in} \sim  10^{17}$~GeV.

The evolution of the lighter stau mass with $M_{in}$ is different in the right
panel of Fig.~\ref{fig:staumass}, where $\lambda = -0.06$. We see that in this case,
after crossing $m_\chi$ and rising further
for low $M_{in} \lappeq 10^{18}$~GeV, the stau mass falls
again for larger $M_{in}$.
Here,  there is a second region where stau-$\chi$ coannihilation is
important and the relic density is brought within the WMAP range: one is just after the
lower crossing point and the other is very close to $M_{in} = \mplr$. 
The base and hypotenuse of the WMAP triangles in panel (c) of Fig.~\ref{fig:m12min1}
and panels (b) and (c) of Fig.~\ref{fig:m12min2} can be understood as these upper
and lower coannihilation strips. Likewise, the doughnut in panel (d) of Fig.~\ref{fig:m12min1}
and the blob in panel (d) of Fig.~\ref{fig:m12min2} can be understood as cases where
the stau mass never rises much above $m_\chi$ so that the two coannihilation strips
merge, and the disappearances of the
doughnut and blob for larger values of $\lambda$ are because the stau mass never rises
(far enough) above $m_\chi$.
The same stau coannihilation mechanism operates in the nearly horizontal parts of Figs.~\ref{fig:positive} and \ref{fig:m12minlrat}.

\begin{figure}
\begin{center}
\epsfig{file=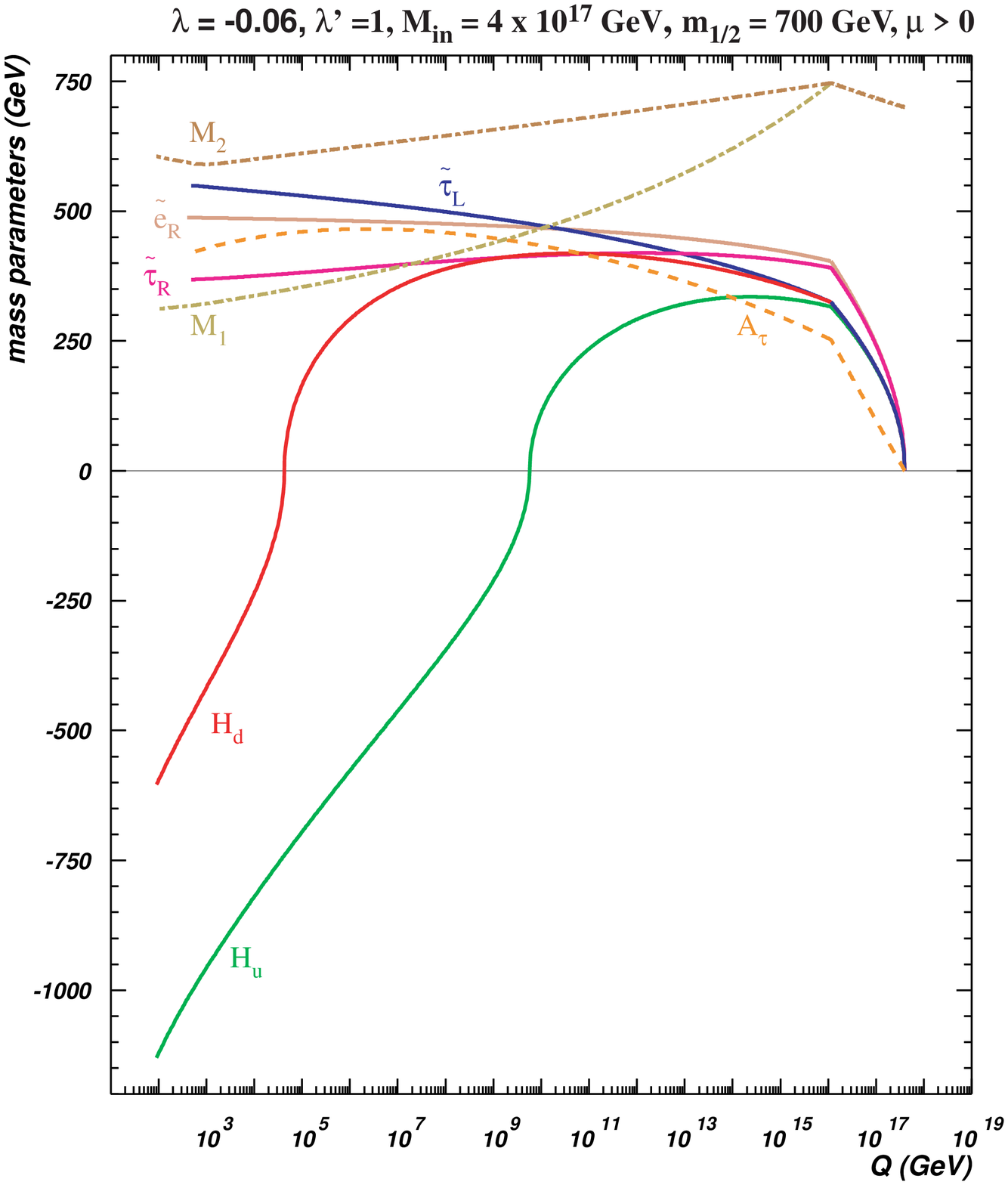,height=8cm}
\epsfig{file=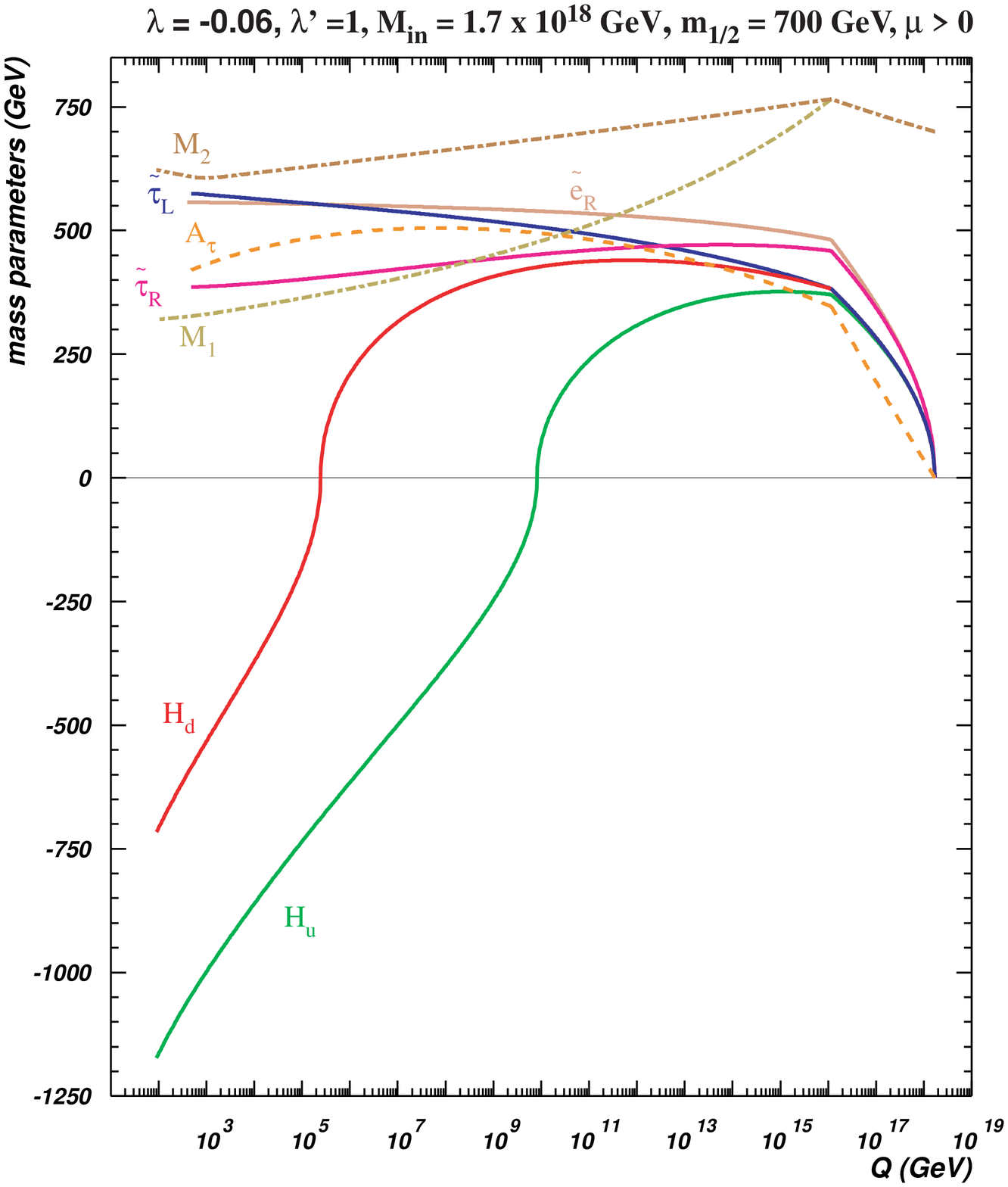,height=8cm}
\end{center}
\caption{\it
The RG evolutions of the supersymmetry-breaking parameters
in the no-scale SU(5) model for $\lambda = -0.06, \lambda' = 1$
and $m_{1/2} = 700$~GeV, as a function of the renormalization scale $Q$,
assuming $M_{in} = 4 \times 10^{17}$~GeV (left) and $1.7 \times 10^{18}$~GeV (right). 
Note the rapid evolution of the trilinear parameter above $M_{GUT}$:
the no-scale boundary condition at $M_{in}$ leads to non-zero $A_\tau$ at $M_{GUT}$.}
\label{fig:evolution}
\end{figure}

These differences in the dependence of $m_{\stau_1}$ may be traced to the influence
of the trilinear supersymmetry-breaking parameter $A_\tau$. Specifically, the
separation between $m_{\tilde e_R}$ and $m_{\stau_1}$ that increases with $M_{in}$,
which is visible in both
panels of Fig.~\ref{fig:staumass}, is due to $A_\tau$. The growth of the separation
with $M_{in}$ reflects the rapid RG evolution of $\afiv$ between $M_{in}$ and the GUT
scale seen in both panels of Fig.~\ref{fig:evolution}. As seen in (\ref{RGEAb}),
this evolution is sensitive
to the value of $\lambda$, but not to $\lambda'$, except indirectly via the RGE
for $\lambda$. In
Fig.~\ref{fig:evolution}, we have assumed $\lambda = -0.06$: the effect of varying
$\lambda$ can be seen by comparing the left and right panels of Fig.~\ref{fig:staumass}.
In the left panel with $\lambda = -0.04$, the separation between $m_{\tilde e_R}$ and 
$m_{\stau_1}$ is much smaller than in the right panel, where $\lambda = -0.06$ has 
been assumed, and the renormalization drives $m_{\stau_1}$ back close to $m_\chi$
at large $M_{in}$.

In the near-vertical part of the WMAP-allowed strips of Figs.~\ref{fig:m12min1} -- \ref{fig:m12minlrat},
$\ohsq$ is lowered by a different mechanism. In this region, the $\stau_1$ is still the next-to-lightest
supersymmetric particle
but, due to the smallness of $m_{1/2}$, the $\stau_1 -\chi$ mass gap is too large for 
stau coannihilation
to be effective. The smallness of $m_{1/2}$ also means that the
soft supersymmetry-breaking scalar masses are not pushed very high during
the SU(5) evolution from $M_{in}$ to $\mgut$, resulting in a lighter sfermion spectrum. 
This in turn enables neutralinos to pair-annihilate efficiently into 
$\tau$ leptons via the exchange of staus in
the $t$-channel -- the same mechanism that operates in the bulk region of 
the CMSSM~\cite{EHNOS,cmssm}.
Again just as in CMSSM, the light sfermion spectrum also leads to a
light Higgs boson, creating tension with the LEP
bound on $m_h$.

We also see in Fig.~\ref{fig:staumass} that the
mass of the CP-odd Higgs boson $A$ decreases as
$M_{in}$ grows. This is because larger values of
$M_{in}$ lead to larger $\tan\beta$, as explained earlier. The larger value of
$\tan\beta$ in turn produces a greater downward push of the $H_d$ mass-squared parameter, 
resulting in a
smaller value of $m_{H_d}^2$ at the weak scale, as can be seen in Fig.~\ref{fig:evolution}.
Therefore the CP-odd Higgs boson mass, that is given at tree level by the expression 
$m_A^2 \simeq m^2_{H_d}-m^2_{H_u}$, decreases as $M_{in}$ increases. 
At some point it approaches $2m_\chi$,
the condition necessary for the resonant pair annihilation via the direct
Higgs channel~\cite{funnel,efgosi}. However, $m_{\stau_1}$ also
decreases as $M_{in}$ increases, and $\stau_1$ always becomes the LSP before the 
resonance condition $m_A \simeq
2m_\chi$ is reached. This can be traced to the dependence of $\tan\beta$ on $m_{1/2}$ 
in the no-scale scenario. Only if this dependence is broken as in Ref.~\cite{EMO,quasinoscale}, 
where $\tan\beta$ is taken as a free parameter, can the $A$-funnel be reconciled
with a framework with $m_0 =A_0 = 0$ and
universal gaugino masses.

\section{Sparticle Detection Prospects}

\begin{figure}[ht]
\begin{center}
\epsfig{file=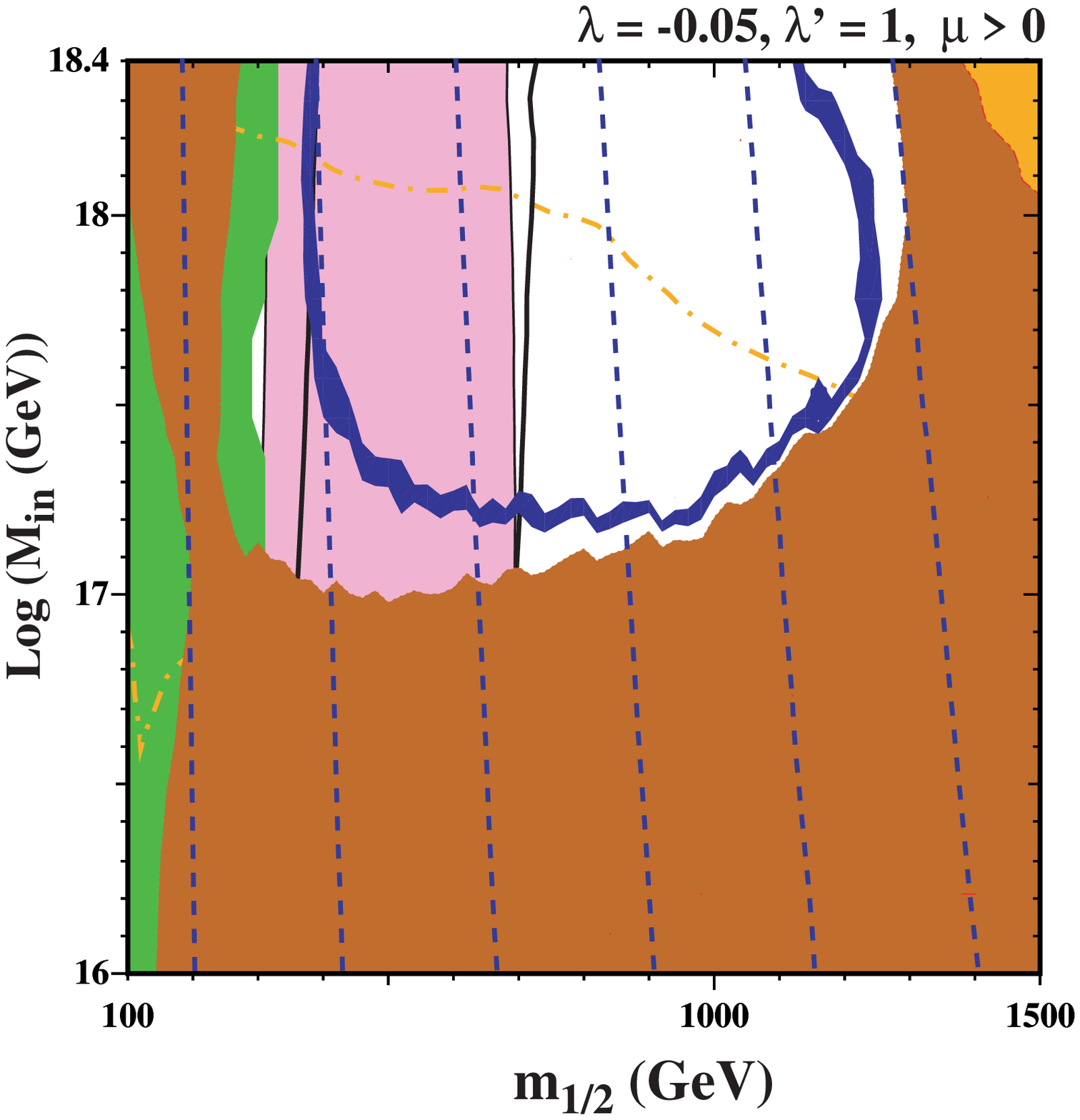,height=8cm}
\epsfig{file=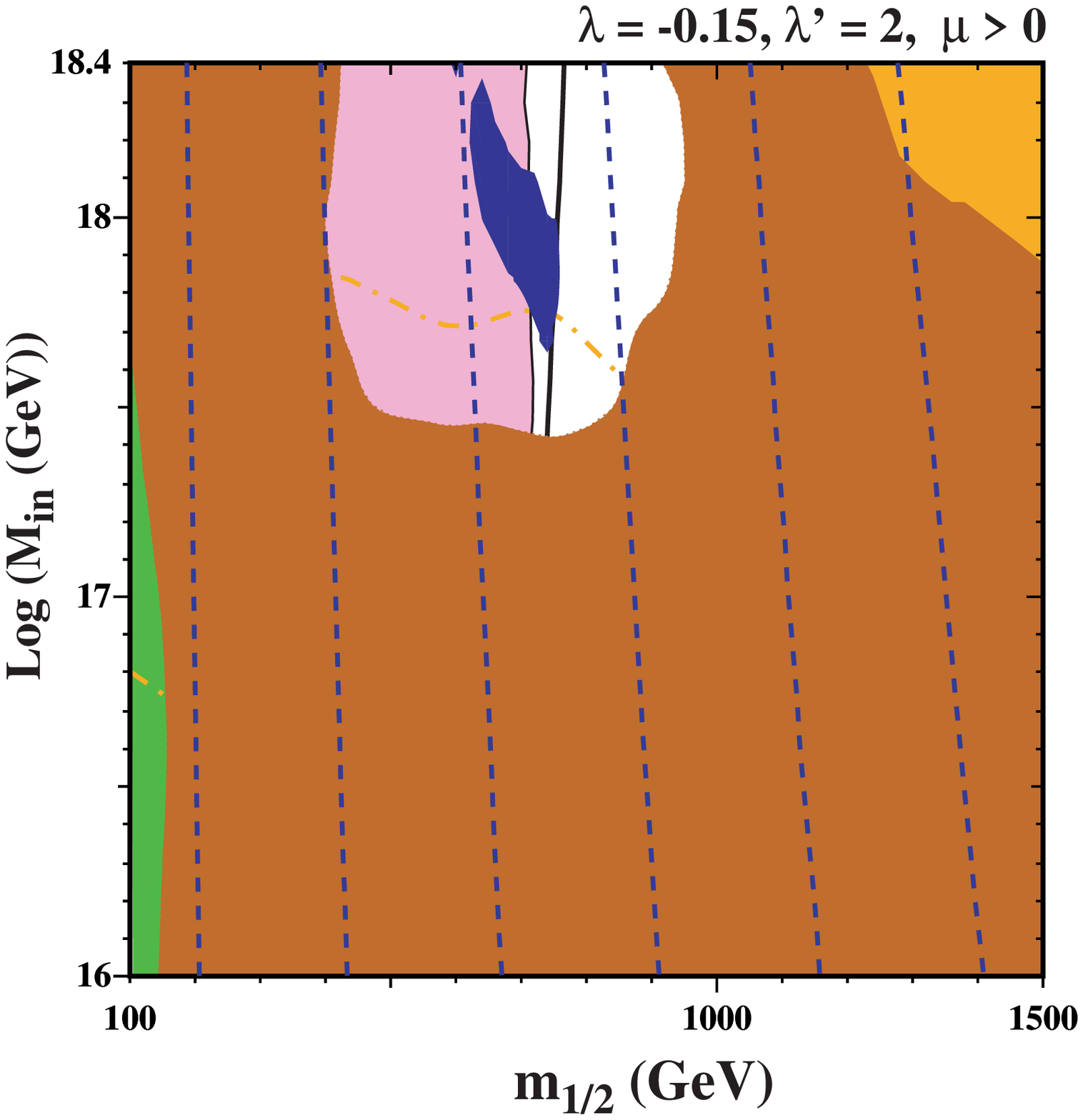,height=8cm}
\end{center}
\caption{\it
The $(m_{1/2}, M_{in})$ planes for $(\lambda, \lambda') = (-0.05, 1)$ and $(-0.15, 2)$ in the left and right plots, respectively, displaying contours of $m_{\tilde g} = 500, 1000,  1500, 2000, 2500$ and $3000$~GeV (blue dashed lines), $\sigma_{SI} = 10^{-6}, 10^{-7}, 10^{-8}$ and $10^{-9}$~pb (black solid lines) and $h_b/h_\tau = 0.65$ and $0.6$ (orange dashed lines). 
The other curves and shadings have the same interpretations as for Fig.~\protect\ref{fig:m12min1}.}
\label{fig:detection}
\end{figure}

We conclude by discussing the observability of sparticles in the no-scale models
discussed here. We see from Figs.~\ref{fig:m12min1} and \ref{fig:m12min2} that
the portions of the WMAP-compatible no-scale strip that are also compatible
with $g_\mu - 2$ at the 2-$\sigma$ level have $250~{\rm GeV} \lappeq m_{1/2}
\lappeq 700$~GeV, compared with $m_{1/2} \lappeq 1500$~GeV if the $g_\mu - 2$
constraint is not applied. Since the only low-energy mass-scale is $m_{1/2}$, the
low-energy spectrum is roughly proportional to $m_{1/2}$. Fig.~\ref{fig:detection} displays
contours of $m_{\tilde g} = 500, 1000, 1500, 2000, 2500$ and $3000$~GeV 
(blue dashed lines) for the choices $(\lambda, \lambda') = (-0.05, 1)$ (left plot)
and $(-0.15, 2)$ (right plot)~\footnote{The locations of these contours are relatively insensitive
to the values of $\lambda$ and $\lambda'$.}. We see that $m_{\tilde g} \lappeq 1700$~GeV
in the areas favoured by $g_\mu - 2$ at the 2-$\sigma$ level, and may range up to
$m_{\tilde g} \sim 2800$~GeV if the $g_\mu - 2$ constraint is discarded, still within
the range of the LHC. We therefore conclude that the
no-scale scenario is in principle testable in the near future.
The spectra shown in Figs.~\ref{fig:staumass}
and \ref{fig:evolution} illustrate the range of possibilities for the sparticle masses
for the case $m_{1/2} = 700$~GeV, which is in the upper part of the possible range for $m_{1/2}$
if the $g_\mu - 2$ constraint is applied.
We see that the heaviest squark may weigh up to $\sim 1350$~GeV, while the
gluino may weigh up to $\sim 1450$~GeV: both of these are well within the
reach of the LHC. Slepton masses range up to $\sim 500$~GeV,
within the reach of a 1-TeV linear collider.

Another approach to testing the no-scale model discussed here is through the 
direct detection of the LSP.  Here, we follow the calculation outlined recently in
\cite{dd} using $\Sigma_{\pi N} = 64$ MeV: for details on the sensitivity to this 
choice, see \cite{dd2}. 
Shown in Fig.~\ref{fig:detection} are contours of the spin-independent LSP dark
matter scattering cross section $\sigma_{SI} =  10^{-8}$ and $10^{-9}$~pb (black solid lines - only the latter is visible in panel (b)). We see that values of $\sigma_{SI} \sim 10^{-8}$~pb are typical, with a range 
extending less than an order of magnitude above and below if the $g_\mu - 2$ constraint is applied,
somewhat more if it is discarded. We also display in Fig.~\ref{fig:detection} contours of
$h_b/h_\tau = 0.65$ and $0.6$ as the lower and upper orange dashed lines, respectively.
We see that these no-scale models violate $b - \tau$ Yukawa unification~\cite{CEG} 
quite significantly, pointing to the need either for higher-order non-renormalizable
Yukawa couplings in the superpotential~\cite{EG} or additional dynamics, e.g., in the neutrino
sector~\cite{CELW}, that might not affect other aspects of the analysis presented here.

In contrast, in the CMSSM the possible range of $m_{1/2}$ is larger,
particularly in the focus-point region~\cite{focus} and when there is rapid LSP-pair
annihilation through the heavy Higgs bosons $H, A$ in the direct channel~\cite{funnel,efgosi}.
In both these regions, $m_0$ also takes large values, an impossibility in
no-scale models. Because of these large values of $m_{1/2}$ and $m_0$,
the discovery of sparticles at the LHC cannot be guaranteed in the CMSSM~\footnote{The
same is true in more general models such as the NUHM~\cite{nuhm}.}.
It was to be expected that the range of experimental possibilities would be
restricted in no-scale models: the good news is that the restriction is to
accessible ranges of sparticle masses.

\section{Acknowledgements}
The work of A.M. and K.A.O. is supported in part by DOE grant DE-FG02-94ER-40823 at the 
University of Minnesota.

\end{document}